\documentclass[review]{elsarticle}
\usepackage{array}
\usepackage{amssymb}
\usepackage{amsmath}
\usepackage{amsthm}
\usepackage{graphicx}
\usepackage{lineno,hyperref}

\def\sumi{\sum_{i=1}^n}
\def\xia{\vx_{A_i}}

\newcommand{\va}{{\bf a}}
\newcommand{\vb}{{\bf b}}

\newcommand{\vl}{{\bf l}}
\newcommand{\vW}{{\bf W}}

\newcommand{\vm}{{\bf m}}

\newcommand{\vt}{{\bf t}}
\newcommand{\vs}{{\bf s}}
\newcommand{\vu}{{\bf u}}

\newcommand{\vw}{{\bf w}}
\newcommand{\vx}{{\bf x}}

\newcommand{\vz}{{\bf z}}

\newcommand{\vX}{{\bf X}}

\newcommand{\vZ}{{\bf Z}}

\newcommand{\vgamma}{\mbox{\boldmath $\gamma$}}
\newcommand{\vbeta}{\mbox{\boldmath $\beta$}}
\newcommand{\vxi}{\mbox{\boldmath $\xi$}}
\newcommand{\veta}{\mbox{\boldmath $\eta$}}
\newcommand{\vdelta}{\mbox{\boldmath $\delta$}}
\newcommand{\vOmega}{\mbox{\boldmath $\Omega$}}
\DeclareMathOperator*{\argmin}{arg\,min}

\newcommand{\vWzi}{\vW(\vz_i)}
\newcommand{\vtWzi}{\tilde{\vW}(\vz_i)}
\newcommand{\vzi}{\vz_i}
\newcommand{\vtheta}{\mbox{\boldmath $\theta$}}

\def\convInD{\overset{d}{\rightarrow}}

\newtheorem{theorem}{Theorem}
\newtheorem{lemma}{Lemma}

\newtheorem{condition}{Condition}
\newenvironment{definition}[1][Definition]{\begin{trivlist}
\item[\hskip \labelsep {\bfseries #1}]}{\end{trivlist}}

\begin{document}


\renewcommand{\baselinestretch}{1.2}

\markboth{\hfill{\footnotesize\rm Ben Sherwood} \hfill}
{\hfill {\footnotesize\rm Variable Selection for Additive Partial Linear Quantile Regression with Missing Covariates} \hfill}

\renewcommand{\thefootnote}{}
$\ $\par


\fontsize{10.95}{14pt plus.8pt minus .6pt}\selectfont
\vspace{0.8pc}
\centerline{\large\bf Variable Selection for Additive Partial Linear Quantile Regression}
\vspace{2pt}
\centerline{\large\bf  with Missing Covariates}
\vspace{.4cm}
\centerline{Ben Sherwood}
\vspace{.4cm}
\centerline{\it Johns Hopkins University}
\vspace{.55cm}
\fontsize{9}{11.5pt plus.8pt minus .6pt}\selectfont


\begin{quotation}
\noindent {\it Abstract:}
The standard quantile regression model assumes a linear relationship at the quantile of interest and that all 
variables are observed. These assumptions are relaxed by considering a partial linear model with missing covariates. A weighted objective function using inverse probability weighting can be used to remove the potential bias caused by missing data. Estimators using parametric and nonparametric estimates of the probability an observation has fully observed covariates are examined. A penalized and weighted objective function using the nonconvex penalties MCP or SCAD is used for variable selection of the linear terms in the presence of missing data. Assuming the missing data problems remains a low dimensional problem the penalized estimator has the oracle property including cases where $p>>n$. Theoretical challenges include handling missing data and partial linear models while working with a nonsmooth loss function and a nonconvex penalty function.  The performance of the method is evaluated using Monte Carlo simulations and the methods are applied to model amount of time sober for patients leaving a rehabilitation center.   
\par

\vspace{9pt}
\par
\end{quotation}\par


\fontsize{10.95}{14pt plus.8pt minus .6pt}\selectfont

\section{Introduction}

Linear quantile regression, proposed by Koenker and Bassett (1978), provides an estimate of a conditional quantile without requiring any distributional assumptions. It assumes a linear relationship between the response and the covariates at the quantile of interest and that all variables are fully observed. This paper introduces an additive partial linear model that can accommodate missing covariates. Inverse probability weighting (IPW) is used to remove potential bias caused by missing data. The IPW framework fits nicely with quantile regression because it does not require any distributional assumptions for the covariates or response. However, it does require a model for the probability an observation has complete data. To provide flexibility the probabilities can be estimated using parametric and nonparametric methods. A weighted and penalized objective function is proposed for variable selection of the linear covariates in the presence of missing data. 

Consider the sample $\{Y_i, \vx_i, \vz_i\}_{i=1}^n$ with $Y_i \in \mathbb{R}$, $\vx_i = (x_{i1},\ldots,x_{ip})^\top \in \mathbb{R}^p$ and $\vz_i = (z_{i1},\ldots,z_{id})^\top \in \mathbb{R}^d$. The conditional quantile of $Y \mid \{\vx, \vz\}$ for a fixed $\tau \in (0,1)$ is defined as $Q_\tau(\vx,\vz)$ where $\Pr(Y \leq Q_\tau(\vx,\vz) \mid \vx,\vz) = \tau$. This paper considers a model where for a given value of $\tau$ $Q_\tau(\cdot)$ has a linear relationship with $\vx$ and an unknown relationship with $\vz$ for the model
\begin{equation}
\label{origModel}
Y_i = \vx_i^\top\vbeta_{0}(\tau) + g_0(\vz_i, \tau) + \epsilon_i,
\end{equation}
where $\Pr(\epsilon_i < 0 \mid \vx_i,\vz_i) = \tau$ and $g_0(Z_i,\tau) = \alpha_{0}(\tau)  + \sum_{j=1}^d g_j(z_{ij},\tau)$. All observable variables, $\left(Y_i,\vx_i,\vz_i\right)$, are assumed to have have a marginal i.i.d., but this still allows for common cases of $\epsilon_i \mid \{\vx_i, \vz_i\}$ that are n.i.d. Recent work in estimating conditional quantiles in the presence of missing data (Chen, Wan and Zhou (2015)) made a major distinction between the i.i.d. and n.i.d. errors. Section 2 includes a discussion of the philosophy on why the marginal i.i.d. assumption is reasonable and can accommodate cases of n.i.d. errors. For model identifiability it is assumed that ${\rm E}\{g_j(z_{ij})\}=0$ $\forall j$ and the intercept is part of the unknown nonlinear function. For technical simplicity it is assumed that ${\rm E}\{\vx_i\}=0$. The additive model for $\vz$ allows for $d$ nonlinear functions while avoiding ``the curse of dimensionality". The intercept $\alpha_0(\tau)$, linear coefficients $\vbeta_0(\tau)$ and nonlinear function $g_0(z, \tau)$ all depend on $\tau$. For ease of notation $\tau$ will be dropped for the rest of the article. 
Without loss of generality, the first $q$ coefficients of $\vbeta_0$ are nonzero and the remaining $p-q$ coefficients are zero. Formally, $\vbeta_0 = (\vbeta_{01}^\top, \mathbf{0}_{p-q}^\top)^\top$ with $\vbeta_{01} \in \mathbb{R}^{q}$ and $\mathbf{0}_{p-q} \in \mathbb{R}^{p-q}$. Let $X = (\vx_1,\ldots,\vx_n)^\top$ be the $n \times p$ design matrix of linear covariates and $X_A = (\vx_{A_1},\ldots,\vx_{A_n})^\top$ be the $n \times q$ submatrix of the active linear covariates corresponding to the first $q$ columns of $X$. The case of fixed and increasing $q$ and $p$ are considered, when considering the increasing case the covariates are indexed as $q_n$ and $p_n$. 

This paper addresses estimating (\ref{origModel}) when a subset of $\{\vx_i,\vz_i\}$ 
has values that are not always observed and the missing at random assumption holds. Robins et al. (1994) proposed handling the potential bias from missing data by using a weighted estimating equation. Estimates are derived using observations with complete data, but the weights account for the missing observations. Weights are assigned by using inverse probability weighting (IPW), that is, the weight for an observation is the inverse of the probability of an observation with the same observed variables having complete data. Thus observations similar to those with missing data will receive larger weights. 

Wang, Wang, Gutierrez and Carroll (1998) consider using the weighted approach with a local linear smoother for a generalized univariate linear model with missing data. Liang et al. (2004) applied the IPW approach to partial linear mean models, but assumed the nonlinear covariates were always observed. Lipsitz et al. (1997) and Yi and He (2009) proposed IPW methods for longitudinal quantile regression models with dropouts. Sherwood et al. (2013) used the IPW approach with linear quantile regression and proposed a BIC type procedure with a weighted objective function for model selection. Liu and Yuan (2015) proposed a weighted empirical likelihood quantile regression estimator for missing covariates that achieves semiparametric efficiency. Wei et al. (2012) presented a multiple imputation solution for linear quantile regression. They assume the linear model holds for all quantiles and address efficiency loss, but not bias, caused by missing covariates. Wei and Yang (2014) use multiple imputation to handle bias caused by missing covariates in linear quantile regression. They assume a location-scale model and limit missing covariates to those that only influence the location. Chen, Wan and Zhou (2015) provide a thorough analysis of efficiency using IPW with linear quantile regression. They propose three different estimating equations using IPW for linear quantile regression with missing covariates. They estimate the weights nonparametrically and demonstrate that the estimators achieve the semiparametric efficiency bound. They propose an estimating equation approach and focused on linear quantile regression, while the methods in this paper work directly with the objective function and relaxes the linearity assumption. 

He and Shi (1996) proposed a partial linear quantile regression model using B-splines that did not assume an additive structure but, limited their model to $d=2$. Partial linear quantile regression has been extended to a variety of settings including longitudinal models (He et al. (2002)), for $d=1$, and varying coefficient models (Wang et al. (2009)), for $d=1$ but $p$ unknown functions are estimated. Horowitz and Lee (2005) and De Gooijer and Zerom (2003) proposed fully nonparametric additive quantile regression models. 

This paper provides several new contributions. First, IPW is extended to the partial linear quantile regression model. Second, a penalized and weighted objective function for simultaneous estimation and variable selection in the presence of missing data is proposed. This paper includes a discussion of why it is reasonable to assume i.i.d. variables in the presence of conditional n.i.d. errors. In addition, it is shown that model selection consistency holds for the high dimensional case of $p >> n$. Techniques used for density estimation, quantile regression, missing data, nonlinear estimation and nonconvex penalties are combined to address these problems. 

First the asymptotic behaviors of the oracle model is analyzed. That is, the partial linear model that only includes the column vectors of $X_A$ as linear covariates. Next, simultaneous estimation and variable selection for all the potential covariates is considered by adding a nonconvex penalty function to the weighted objective function. In this work it is assumed that the nonlinear terms are known a priori to be part of the true model and restrict variable selection to the linear terms. However, for the data analysis in Section 5 a weighted BIC approach is proposed to designate variables as linear or nonlinear. The penalized objective function uses nonconvex penalties, either the SCAD penalty (Fan and Li (2001)) or MCP (Zhang (2010)), and under standard conditions the penalized estimator has the oracle property. That is, in the set of local minimums of the nonconvex objective function there exists an estimator that is asymptotically as efficient as if the true linear covariates where known a priori. Liu, et al. (2011) used the SCAD penalty to select the linear components from an additive partial linear mean regression model. Wu and Liu (2009) demonstrated that for linear quantile regression the estimator minimizing the penalized objective function with the SCAD penalty has the oracle property. The use of nonconvex penalties for $p >> n$ has been explored for linear (Wang et al. (2012)) and additive partial linear (Sherwood and Wang (2014)) quantile regression. 

The article is structured as follows. In Section 2 the additive partial linear quantile regression model with missing linear covariates is introduced. Section 3 focuses on a weighted and penalized objective function for model selection of linear covariates in the presence of missing data. The finite sample size performance of the penalized estimator is analyzed using Monte Carlo Simulations in Section 4. In Section 5 the proposed methods are used to model amount of time sober for patients from a rehabilitation center. In addition, Section 5 includes a proposal for a BIC type procedure to designate whether a variable is linear or nonlinear. Section 6 concludes the paper with a summary and discussion of directions for future research.     

\par
\bigskip

\section{Inverse Probability Weighting }
\subsection{Additive partial linear quantile regression}

This section focuses on estimation of (\ref{origModel}) assuming that it is known which linear covariates should be included in the model. This will be insightful for understanding the performance of the method for the low dimensional case that does not require model selection. Understanding the asymptotics for this model is also important for the high dimensional case because in Section 3 the oracle model is shown with probability one to be a local minimizer of the nonconvex penalized objective function. 

Adapting the work of Stone (1985) for the partial linear quantile regression setting, B-splines are used to estimate the additive nonlinear function $g(\cdot)$. Theoretical results for  B-splines assume a compact support and without loss of generality it is assumed that $z_{ij} \in [0,1], \,\,\, \forall i,\ j$. First, two definitions are provided to define the class of functions that can be estimated with B-splines. 
\begin{definition}
Let $r \equiv m + v$. Define $\mathcal{H}_r$ as the collection of functions on $[0,1]$ whose $m$th derivative satisfy the H\"{o}lder condition of order $v$. That is, for any $h \in \mathcal{H}_r$, there exist some positive constant $C$
such that 
\begin{equation}
\label{holder_cond}
\left| h^{(m)}(z') - h^{(m)}(z)\right| \leq C \left|z' - z\right|^v, \,\,\, \forall \ \  0 \leq z', z \leq 1.
\end{equation}
\end{definition}

\begin{definition}
Given $\vz = (z_1,\ldots,z_d)^\top$, the function $g(\vz)$ belongs to the class of nonlinear functions $\mathcal{G}_r$ if $g(\vz) = \sum_{j=1}^d g_j(z_j)$,
$g_j \in \mathcal{H}_r$ and ${\rm E}\{g_j(z_{j})\}=0 \, \forall j$.
\end{definition}
If $g_0 \in \mathcal{G}_r$ for some $r > 1.5$ then $g_0$ can be approximated using B-spline basis 
functions. To construct the B-splines the unit interval is partitioned into $k_n$ subintervals such that $0 = s_0 < \ldots < s_{k_n} = 1$.  Then $s_i, \, i = 0, \ldots , k_n$, quasi-uniform knots are used to construct $L_n = k_n + l$ normalized B-spline basis functions of degree $l$. 
For a given covariate $z_{ik}$, let $\tilde{b}(z_{ik})= \left(b_0(z_{ik}),\ldots,b_{L_n}(z_{ik})\right)^\top$ denote 
the corresponding vector of B-spline basis functions of degree $l$, where the $l$ index is dropped for ease of notation. A property of B-splines is that $\sum_{j=0}^{L_n}b_j(z_{ik}) = 1$, thus to avoid collinearity when fitting models only $b(z_{ik})= \left(b_1(z_{ik}),\ldots,b_{L_n}(z_{ik})\right)^\top$ is used. Define $\vb(\vz_i)=\left(1,b(z_{i1})^\top,\ldots,b(z_{id})^\top\right)^\top \in \mathbb{R}^{dL_n+1}$ and $B = (\vb(\vz_1),\ldots,\vb(\vz_n))^\top \in \mathbb{R}^{n \times dL_n + 1}$ as the matrix of the basis transformations.
The constant term is included because for identifiability purposes the intercept is considered part of the unknown function $g_0(\cdot)$. For ease of notation and proofs the same number of internal knots is used for each nonlinear variable, but in application the number of knots can vary between the $d$ variables. 

One benefit of B-splines is estimation of the linear and nonlinear components can be done in a single step. Define $\rho_\tau(u) = u(\tau-I(u < 0))$. If all covariates are observed consistent estimates can be obtained by
\begin{equation}
\label{completeObj}
\left(\hat{\vbeta}_1, \hat{\vxi}\right) = \underset{\vbeta_1, \vxi}{\argmin} \sumi \rho_\tau(Y_i - \vx_{A_i}^\top\vbeta_1 - \vb(\vz_i)^\top\vxi),
\end{equation}
where $\vxi \in \mathbb{R}^{dL_n+1}$ is the vector of coefficients for the intercept and B-spline basis of the nonlinear covariates. The estimator of the nonparametric function is defined as $\hat{g}(\vz_i) = \vb(\vz_i)^\top\hat{\vxi}$. 

\par
\bigskip
\subsection{Quantile regression with missing covariates}

Next, estimation of (\ref{origModel}) is considered when a subset of the covariates have missing values. Let $\vl_i \in \mathbb{R}^{p+d-k}$ be a vector of always observed covariates and $\vm_i \in \mathbb{R}^k$ be a vector of covariates that may contain some missing components. While the oracle model has only $q$ covariates we consider all $p$ covariates for the missing data setup and covariates in $\vm_i$ and $\vl_i$ can be from $\vx_i$ or $\vz_i$. In other words, no relationship is assumed between the missingness and whether a covariate has a linear or nonlinear relationship with the response or missingness and whether a covariate is part of the true model. For each observation, an indicator variable $R_i$ denotes if $\vm_i$ is fully observed, that is, $R_i=1$ if $\vm_i$ is fully observed, and $R_i=0$ otherwise. Let $\vt_i = (Y_i, \vl_i^\top)^\top \in \mathbb{R}^{s}$, with $s=p+d-k+1$, which is a vector of variables that are always observed. The data is assumed to be missing at random (MAR). That is, the probability an observation is missing can depend on variables that are always observed, but not variables that may have missing data. Formally, 
\begin{equation*}
\Pr(R_i = 1 \mid Y_i, \vx_i, \vz_i) = \Pr(R_i = 1 \mid \vt_i) \equiv \pi_0(\vt_i) \equiv \pi_{i0}.
\end{equation*}

A naive approach of estimating (\ref{origModel}) in the presence missing covariates is to fit the model using
only observations with complete data.  The naive estimator is
\begin{equation}
\label{naive}
\left(\hat{\vbeta}_1^N,\hat{\vxi}^N\right) = \underset{(\vbeta_1,\vxi)}{\operatorname{argmin}} \sum_{i=1}^n R_i \rho_\tau (Y_i - \vx_{A_i}^\top\vbeta_1 - \vb(\vz_i)^\top\vxi),
\end{equation}
which estimates the model by dropping all observations with missing data. Under the MAR assumption this estimator will be asymptotically biased. 

An objective function with inverse probability weights (IPW) is proposed to alleviate potential bias caused by missing data. The IPW method weights the $i$th data point by $R_i/\pi_{i0}$. IPW differs from the naive method by providing different weights to records with fully observed data. The intuition behind weighting is that for every fully observed data point with probability $\pi_{i0}$ of being fully observed, $1/\pi_{i0}$ data points with the same covariates are expected if there was no missing data. For example, an observation with complete data and $\pi_{i0}=.25$ is given the weight of four observations. This is to account for the three observations with similar covariates that are likely to have incomplete data.

The weight $1/\pi_{i0}$ is often unknown and needs to be estimated. Estimating the weights using a parametric and nonparametric model are both considered. The parametric model assumes a general parametric relationship of 
\begin{equation*}
\pi_{i0} \equiv \pi_i(\vt_i, \veta_0).
\end{equation*}
One example would be assuming the logistic relationship of
\begin{equation*}
\pi_i(\vt_i, \veta_0) = \frac{e^{\left([1,\vt_i]^\top\veta_0\right)}}{1+e^{\left([1,\vt_i]^\top\veta_0\right)}}.
\end{equation*} 
In practice $\pi_i(\vt_i, \veta_0)$ is replaced with $\pi_i(\vt_i,\hat{\veta}) \equiv \pi_i(\hat{\veta})$ where $\hat{\veta}$ has been estimated using the parametric model for $\Pr(R_i = 1 \mid \vt_i)$. 

The nonparametric approach follows the work of Wang et al. (1997) for linear mean regression and Chen, Wan and Zhou (2015) for linear quantile regression by using a kernel smoother (Nadaraya (1964); Watson (1964)) as an estimator of $\pi_{i0}$. The nonparametric estimator of $\pi_{i0}$ is defined as 
\begin{equation}
\label{kern}
\tilde{\pi}_i = \frac{\sum_{j=1}^n R_jK_h(\vt_i-\vt_j)}{\sum_{j=1}^n K_h(\vt_i-\vt_j)}.
\end{equation}
Where $K_h(\nu) = K(\nu/h)/h^s$, is a $s$-variate kernel function and $h$ is the bandwidth variable. One could have  different bandwidths, $h_1, \ldots, h_s$, for the $s$ variables, but for simplicity one bandwidth variable is used. Throughout the paper $\pi_i(\hat{\veta})$ will denote the parametric estimate, $\tilde{\pi}_i$ will denote the kernel based estimate, $\hat{\pi}_i$ will denote a general estimate that could be kernel based or parametric and $\pi_{i0}$ will denote the true probability observation $i$ has complete data. The parametric weighted quantile regression estimator is defined as

\begin{equation}
\label{weighted_p}
\left(\hat{\vbeta}_1^{P},\hat{\vxi}^{P}\right) = \underset{\vbeta_1,\vxi}{\operatorname{argmin}} \sum_{i=1}^n \frac{R_i}{\pi_i(\hat{\veta})} \rho_\tau (Y_i - \vx_{A_i}^\top\vbeta_1 - \vb(\vz_i)^\top\vxi),
\end{equation}
with the estimator of the nonparametric function defined as $\hat{g}^{P}(\vz_i) = \vb(\vz_i)^\top\hat{\vxi}^{P}$. Liang et al. (2004) considered a similar model for mean regression, but used a local linear kernel method to estimate the nonlinear terms and assumed all nonlinear covariates were observed. The nonparametric weighted quantile regression estimator is defined as 
\begin{equation}
\label{weighted_n}
\left(\hat{\vbeta}_1^{K},\hat{\vxi}^{K}\right) = \underset{\vbeta_1,\vxi}{\operatorname{argmin}} \sum_{i=1}^n \frac{R_i}{\tilde{\pi}_i} \rho_\tau (Y_i - \vx_{A_i}^\top\vbeta_1 - \vb(\vz_i)^\top\vxi),
\end{equation}
with the estimator of the nonparametric function defined as $\hat{g}^{K}(\vz_i) = \vb(\vz_i)^\top\hat{\vxi}^{K}$.

\par
\bigskip
\subsection{Asymptotic Results}

To understand the asymptotic behavior of coefficients for the linear terms requires formally defining a relationship between $X$ and $Z$.
Define the set $\mathcal{H}^{d}_r = \big\{\sum_{k=1}^{d} h_k(z) \mid h_j \in \mathcal{H}_r\big\}$ and 
\begin{equation*}
h^*_j(\cdot) = \underset{h_j \in \mathcal{H}_r^d}{\mbox{arg inf}} \sumi {\rm E}\left\{ f_i(0 \mid \vx_i,\vz_i) (x_{ij} - h_j(\vz_i))^2 \right\},
\end{equation*}
where $f_i(\cdot \mid \vx_i,\vz_i)$ is the conditional probability density function of $\epsilon_i \mid \{\vx_i,\vz_i\}$ with $F_i(\cdot \mid \vx_i, \vz_i)$ representing the corresponding conditional cumulative distribution function.
Let $k_j(z) = {\rm E}\left\{x_{ij} \mid \vz_i\right\}$, then $h^*_j$ is the weighted projection of 
$k_j(\cdot)$ into $\mathcal{H}_r^d$ under the $L_2$ norm, where the weights $f_i(0 \mid \vx_i,\vz_i)$ are included to account for possibly heterogeneous errors. 
Furthermore, let $x_{ij}$ be the element of $X$ at the $i$th row and the $j$th column. Define $\delta_{ij} \equiv x_{ij} - h^*_j(\vz_i)$, $\vdelta_{i} = \left(\delta_{i1},\ldots,\delta_{ip}\right)^\top \in \mathbb{R}^{p}$,
$i=1,\ldots, n$ and $\Delta_n = \left(\vdelta_1,\ldots,\vdelta_n\right)^\top \in \mathbb{R}^{n \times p}$.
Define $H$ as the $n \times p$ matrix with the $(i,j)$th element $H_{ij}= h_j^*(\vz_i)$. Then $X = H + \Delta_n$, where $\Delta_n$ is the combined error and bias from estimating $X$ with a function of $Z$ in the set $\mathcal{H}_r^d$ and will be an important part of the asymptotic variance of $\hat{\vbeta}_1^{P}$ and $\hat{\vbeta}_1^{N}$. Wang et al. (2009) used a similar setup to characterize the asymptotic distribution of the parametric components of a quantile regression varying coefficient model.  \\

Asymptotic results are established using the following standard conditions. 

\begin{condition}
\label{cond_f}
(Conditions on the random error)
The conditional probability density function of $f_i(\cdot \mid \vx_i, \vz_i)$ is uniformly bounded away from 0 and
infinity in a neighborhood of zero, its first derivative $f_i'(\cdot \mid \vx_i,\vz_i)$ has a uniform upper bound in a neighborhood of zero,
for $1\leq i\leq n$. 
\end{condition}

\begin{condition}
\label{cond_x}
(Conditions on the observed variables)
There exist a positive constant $M_1$ such that $|x_{ij}| \leq M_1$, $\forall \ 1\leq i \leq n, \, 1 \leq j \leq p$. For the nonlinear covariates $z_{ij} \in [0,1] \,\,\, \forall \ 1\leq i \leq n, \, 1 \leq j \leq d$. In addition, all observable variables $(Y_i, \vx_i, \vz_i)$ have an independent and identical marginal distribution. 
\end{condition}

\begin{condition}
\label{cond_g_h}
(Condition on the nonlinear functions)
For $r=m+v> 1.5$ $g_0 \in \mathcal{G}_r$ and $\forall \ j$, $h_j \in \mathcal{H}_r^d$ and the dimension of the 
internal knots is $k_n \equiv n^{1/(2r+1)}$.
\end{condition}



\begin{condition}
\label{xmar_cond_alpha}
(Condition on the missing probability)
There exists $\alpha_l >0$ and $\alpha_u < 1$ such that $\alpha_l < \pi_{i0} < \alpha_u \, \forall i$.
\end{condition}

\begin{condition}
\label{xmar_cond_eta}
(Condition on parametric estimator of weights)
Assume a parametric form for $\pi_{i0}$ with $\pi_{i0} \equiv \pi_i(\veta_0)$, $\hat{\pi}_i \equiv \pi_i(\hat{\veta})$ and $\hat{\veta}$ is the MLE of:
\begin{align*}
\prod_{i=1}^n \pi_i(\veta)^{R_i} (1-\pi_i(\veta))^{(1-R_i)}.
\end{align*}
With conditions of asymptotic normality of $\hat{\veta}$ holding and $\left|\left| \partial \pi_i(\veta)/\partial \veta\right| \right|$ and
$\left|\left| \partial^2 \pi_i(\veta)/\partial \veta \partial \veta^\top \right| \right|$ are bounded in a neighborhood of $\veta_0$. 
\end{condition}

\begin{condition}
\label{xmar_cond_np}
(Condition on kernel smoothing)
$K(\cdot)$ is an order $b$ kernel function with compact support that satisfies the Lipschitz condition. Let $f_{\vt}(\vt)$, the pdf of $\vt$ that is bounded above zero and below infinity on the support of $\vt$, and $\pi_i(\vt)$ have bounded partial derivatives with respect to $\vt$ up to order $b$. Let $b \geq 2$, $h \rightarrow 0$, $nh^{2s} \rightarrow \infty$, $nh^{2b} \rightarrow 0$ and $nh^{s}/ \ln(n) \rightarrow \infty$.  
\end{condition}

Similar assumptions to Condition \ref{cond_f} and the bounded portion of Conditon \ref{cond_x} are presented in Koenker (2005) for the asymptotic normality of $\hat{\vbeta}_1$ for linear quantile regression with complete data. In addition Condition \ref{cond_x} assumes that all observable variables have a marginal i.i.d. This assumption is discussed in detail after Theorems 1 and 2. Condition \ref{cond_g_h} is a smoothness condition that allows $g_0(\cdot)$ to be well approximated by B-splines. 
The rate of $k_n$ from Condition \ref{cond_g_h} is necessary for optimal rate of convergence of $\hat{g}^W(\cdot)$ similar to results from Stone (1985). The upper bound from Condition \ref{xmar_cond_alpha} is necessary for there to be missing data while the lower bound guarantees that weights do not go to infinity as the sample size increases. Condition \ref{xmar_cond_eta} provides that the parametric weights are asymptotically consistent and is used to understand the asymptotic variance of the weights. Condition \ref{xmar_cond_np} are standard conditons for kernel smoothing estimators and similar to those used by Chen, Wan and Zhou (2015) and Wang et al. (1997).

Define $\psi_\tau(u) = \tau - I(u < 0)$ and 
\begin{eqnarray*}
{\rm E}\left\{ f_i(0 \mid \vx_i, \vz_i) \vdelta_i \vdelta_i^\top\right\} = \Sigma_1, && {\rm E}\left\{\delta_i \frac{1}{\pi_i(\veta_0)} \left( \frac{\partial \pi_i(\veta)}{\partial \veta}\right)^\top_{\veta=\veta_0} \psi_\tau(\epsilon_i)\right\} = \Sigma_3,  \\
{\rm E}\left\{\frac{\psi_\tau(\epsilon_i)^2}{\pi_i(\veta_0)} \vdelta_i \vdelta_i^\top\right\} = \Sigma_2,
&&
{\rm E}\left\{\left( \frac{\partial \pi_i(\veta)}{\partial \veta}\right)_{\veta=\veta_0} \left(\frac{\partial \pi_i(\veta)}{\partial \veta}\right)^\top_{\veta=\veta_0} \frac{1}{\pi_i(\veta_0)(1-\pi_i(\veta_0))}\right\} = I(\veta_0).
\end{eqnarray*} 

Let $\convInD$ denote convergence in distribution. The first theorem is for the estimators $\hat{\vbeta}^{P}$ and $\hat{g}(\cdot)^{P}$. 
\begin{theorem}
\label{xmar_part_lin_asy_normal_p}
Let $\Sigma_m = \Sigma_2 - \Sigma_3I(\veta_0)^{-1}\Sigma_3^\top$. If Conditions \ref{cond_f}-\ref{xmar_cond_eta} hold then 
\begin{eqnarray*}
\sqrt{n}(\hat{\vbeta}_1^{P} - \vbeta_{01}) &\convInD& \mathcal{N}(\mathbf{0},\Sigma_1^{-1}\Sigma_m\Sigma_1^{-1}), \\
\frac{1}{n} \sumi (\hat{g}^{P}(\vz_i) - g_0(\vz_i))^2 &=& {\rm O}_{\rm p}\left(n^{-2/(2r+1)}\right).
\end{eqnarray*}
\end{theorem}

The next theorem is for the estimators $\hat{\vbeta}^{K}$ and $\hat{g}(\cdot)^{K}$. Define 
\begin{equation*}
\Sigma_4 = {\rm E}\left\{ \frac{1-\pi(\vt_i)}{\pi(\vt_i)} {\rm E}\left\{ \vdelta_i \psi_\tau(\epsilon_i) \mid \vt_i\right\}{\rm E}\left\{ \vdelta_i^\top \psi_\tau(\epsilon_i) \mid \vt_i\right\} \right\}.
\end{equation*}

\begin{theorem}
\label{xmar_part_lin_asy_normal_n}
Let $\tilde{\Sigma}_m = \Sigma_2 - \Sigma_4$. If Conditions \ref{cond_f}-\ref{xmar_cond_alpha} and \ref{xmar_cond_np} hold then 
\begin{eqnarray*}
\sqrt{n}(\hat{\vbeta}_1^{K} - \vbeta_{01}) &\convInD& \mathcal{N}(\mathbf{0},\Sigma_1^{-1}\tilde{\Sigma}_m\Sigma_1^{-1}), \\
\frac{1}{n} \sumi (\hat{g}^{K}(\vz_i) - g_0(\vz_i))^2 &=& {\rm O}_{\rm p}\left(n^{-2/(2r+1)}\right).
\end{eqnarray*}
\end{theorem}

\noindent {\bf Remark 1:}
In both theorems estimator of $g_0(\cdot)$ achieves the optimal rate of convergence for an additive function provided by Stone (1985). If the true weights are used the asymptotic distribution of the estimator of $\vbeta_0$ changes to $\mathcal{N}(\mathbf{0},\Sigma_1^{-1}\Sigma_2\Sigma_1^{-1})$. Implying that estimating the weights reduces the variance of an estimator. This is a common result for IPW methods and Robins et al. (1994) discuss the intuition behind this result. It is important to note that the results from Theorem \ref{xmar_part_lin_asy_normal_p} rely on a correctly specified function of the missing probabilities, while Theorem \ref{xmar_part_lin_asy_normal_n} relies on Condition \ref{xmar_cond_np} which places restrictions on the size of $s$, the dimension of the variables influencing the missing probability. 

\

\noindent {\bf Remark 2:}
For $p \times p$ matrices $A$ and $B$ define $A \geq B$ and $A > B$ if $A-B$ is semipositive definite or positive definite respectively. For vectors $\va$ and $\vb$ ${\rm E}\left\{\va\va^\top\right\} \geq {\rm E}\left\{\va\vb^\top\right\}\left({\rm E}\left\{\vb\vb^\top\right\}\right)^{-1}{\rm E}\left\{\vb\va^\top\right\}$ (Tripathi 1999). Define $\va={\rm E}\left\{\vdelta_i\psi_\tau(\epsilon_i)\sqrt{(1-\pi(\vt_i))/\pi(\vt_i)} \mid \vt_i\right\}$ and $\vb=\left(\partial \pi_i(\veta)/\partial \veta\right)_{\veta=\veta_0} \left(\pi(\vt_i)(1-\pi(\vt_i))\right)^{-1/2}$. Then applying the results from Tripathi (1999) $\Sigma_3I(\veta_0)^{-1}\Sigma_3^\top \leq \Sigma_4$ and thus $\tilde{\Sigma}_m \leq \Sigma_m$. Therefore even when the parametric model holds the estimator $\hat{\vbeta}_1^{K}$ is asymptotically at least as efficient as $\hat{\vbeta}_1^{P}$ with equality holding only if $\va = {\rm E}\left\{\va\vb^\top\right\}\left({\rm E}\left\{\vb\vb^\top\right\}\right)^{-1}\vb$ (Lavergne (2008)). This result is similar to results found when using IPW to handle missing covariates with partial linear mean models (Liang, Wang, Robins and Caroll (2004)) and linear quantile regression models (Chen, Wan and Zhou (2015)).

%

\

\noindent {\bf Remark 3:} 
Chen, Wan and Zhou (2015) made distinctions between the i.i.d. and n.i.d. errors because the kernel smoothing methods require $Y_i$ to be i.i.d. They were concerned about this assumption holding when $\epsilon_i$ are n.i.d. This is in contrast to Condition \ref{cond_x} which directly assumes $Y_i$ is i.i.d. The following outlines why this assumption can hold for many cases of n.i.d. errors such as the location scale model. Consider the location scale model 
\begin{equation}
\label{loc_scale}
Y_i = \alpha_{c0} + \vx_i^\top\vbeta_{c0} + \sum_{j=1}^d g_{jc}(z_{ij}) + \left(\alpha_{s0}+\vx_i^\top\vbeta_{s0}+\sum_{j=1}^d g_{js}(z_{ij})\right)\omega_i,
\end{equation}
where $\alpha_{s0}$, $g_{js}(z_{ij})$ and the elements of $\vx_i$ and $\beta_{s0}$ are all non-negative. In addition $\omega_i$ is an i.i.d. random variable from distribution $F$ with quantile function $F_\omega^{-1}(\tau)$, that is $\Pr(\omega \leq F_\omega^{-1}(\tau)) = \tau$. Then for any $\tau \in (0,1)$ model (\ref{loc_scale}) generates the quantile regression model presented in equation (\ref{origModel}) with $\alpha_0(\tau) = \alpha_{c0}+\alpha_{s0}F_\omega^{-1}(\tau)$, $\vbeta_0(\tau) = \vbeta_{c0} + \vbeta_{s0}F_\omega^{-1}(\tau)$ and $g_{j}(z,\tau) = g_{jc}(z) + g_{js}(z)F_\omega^{-1}(\tau)$. In this model the distribution of the error terms are conditionally n.i.d. Let $f_{Y}(y \mid \vX=\vx, \vZ=\vz)$ be the conditional distribution of $Y$ and $f_{\vX,\vZ}(\vx,\vz)$ be the joint distribution of $(\vX,\vZ)$ from model (\ref{loc_scale}). Then the marginal distribution of $Y$ is
\begin{equation}
\label{marginal_y}
f_{Y}(y) = \int_{\vx,\vz} f_{Y}(y \mid \vX=\vx, \vZ=\vz) f_{\vX,\vZ}(\vx,\vz) d\vz d\vz.
\end{equation}
The above marginal distribution is identical for all $Y_i$ even with error terms that are conditionally n.i.d. because the samples are independent and for any $i\neq j$ $f_{Y_i}(y \mid \vX=\vx, \vZ=\vz)=f_{Y_j}(y \mid \vX=\vx, \vZ=\vz)$ for any $\{\vx,\vz\}$. That is the conditional distribution of $Y_i$ and $Y_j$ remains the same if the covariates are the same.  Thus the location-scale setting allows for conditionally n.i.d. error terms and marginal i.i.d. responses. Results presented in this paper would hold if the location-scale model replaced the i.i.d. assumption. The results would also hold under the weaker assumption that for any $i\neq j$ $f_{Y_i}(y \mid \vX=\vx, \vZ=\vz)=f_{Y_j}(y \mid \vX=\vx, \vZ=\vz)$ for any $\{\vx,\vz\}$.

\par
\bigskip

\subsection{High Dimensional Asymptotics}

Under stricter conditions the model can be extended to the high-dimensional case. An important issue is the role of missing data in the high-dimensional data. To avoid all observations having missing data the dimension of the covariates with missing data $\vm_i \in \mathbb{R}^k$ is assumed to be fixed. To simplify the modeling of the missing mechanism the dimension of the covariates involved in the missing model $\vl_i \in \mathbb{R}^{s-1}$ also remain fixed. Thus there are $p_n+d-(k+s-1)$ covariates that are not part of the missing data model and these covariates can be high-dimensional, but the covariates involved in the missing model remain low dimensional.

The most important and restrictive assumption is that the missing data problem remains a fixed dimensional problem. A potential example of this would be a data set with genomic and clinical covariates. The clinical covariates would have missing data with the missingness depending on fully observed clinical covariates. The genomics variable would be high-dimensional, fully observed and have no relationship with the missingness.

The next section examines a nonconvex penalized objective function for simultaneous estimation and model selection. New conditions are required for the high-dimensional setting, both of which are weak for the high-dimensional setting.

\begin{condition}
\label{highd_cond_x}
(Conditions on the covariates)
There exist positive constants $M_1$ and $M_2$ such that $|x_{ij}| \leq M_1$, $\forall \ 1\leq i \leq n, \, 1 \leq j \leq p_n$ and
${\rm E}[\delta_{ij}^{4}] \leq M_2$, $\forall \ 1\leq i \leq n, \, 1 \leq j \leq q_n$.
There exist finite positive constants $C_1$ and $C_2$ such that with probability one
\begin{equation*}
C_1 \leq \lambda_{\max} \left( n^{-1}X_AX_A^\top \right) \leq C_2, \quad
C_1 \leq \lambda_{\max} \left(n^{-1}\Delta_n \Delta_n^\top \right) \leq C_2. 
\end{equation*}
\end{condition}

\begin{condition}
\label{cond_sigma_large_p}
(Condition on model size)
$q_n = O\left(n^{C_3}\right)$ for some $C_3< 1/3$.
\end{condition}

The following theorem states the asymptotics for the estimators from (\ref{weighted_p}) for the high-dimensional setting. 
\begin{theorem}
\label{o_high_d_p}
If Conditions \ref{cond_f}, \ref{cond_g_h}-\ref{xmar_cond_eta} and \ref{highd_cond_x}-\ref{cond_sigma_large_p} hold and $||\hat{\veta}-\veta_0|| = {\rm O}_{\rm p}\left(n^{-1/2}\right)$ then 
\begin{eqnarray*}
\left|\left|\hat{\vbeta}_1^{P} - \vbeta_{01}\right|\right|_2&=& {\rm O}_{\rm p}\left(\sqrt{\frac{q_n}{n}}\right), \\
\frac{1}{n} \sumi (\hat{g}^{P}(\vz_i) - g_0(\vz_i))^2 &=& {\rm O}_{\rm p}\left(\frac{q_n}{n}+n^{-2/(2r+1)}\right).
\end{eqnarray*}
\end{theorem}

The next theorem presents the asymptotics for the estimators from (\ref{weighted_n}) for the high-dimensional setting. 
\begin{theorem}
\label{o_high_d_n}
If Conditions \ref{cond_f}, \ref{cond_g_h}-\ref{xmar_cond_alpha} and \ref{xmar_cond_np}-\ref{cond_sigma_large_p} hold, with pointwise convergence rate of $\left| \tilde{\pi}_i-\pi_{i0} \right| = {\rm O}_{\rm p}\left(h^b+(h^sn)^{-1/2}\right)$ and uniform convergence rate of $\underset{i}{\max} \left| \tilde{\pi}_i-\pi_{i0} \right| = {\rm O}_{\rm p}\left(h^b+(\ln n/h^sn)^{1/2}\right)$ then 
\begin{eqnarray*}
\left|\left|\hat{\vbeta}_1^{K} - \vbeta_{01}\right|\right|_2&=& {\rm O}_{\rm p}\left(\sqrt{\frac{q_n}{n}}\right), \\
\frac{1}{n} \sumi (\hat{g}^{K}(\vz_i) - g_0(\vz_i))^2 &=& {\rm O}_{\rm p}\left(\frac{q_n}{n}+n^{-2/(2r+1)}\right).
\end{eqnarray*}
\end{theorem}

{\bf Remark 1:} Theorems \ref{o_high_d_p} and \ref{o_high_d_n} are generalizations of Theorems \ref{xmar_part_lin_asy_normal_p} and \ref{xmar_part_lin_asy_normal_n}. The major restriction is that modeling of the missing probability remains a fixed dimensional problem. The assumptions of $||\hat{\veta}-\veta_0|| = {\rm O}_{\rm p}\left(n^{-1/2}\right)$ and the pointwise and uniform convergence rates for the kernel estimator are not mentioned in Theorems 1 and 2 because they are standard rates of convergence for the fixed dimensions case, see Theorem 2.2 of Cheng (1995) for the uniform nonparametric rate of convergence. However, for the high-dimensional case they are much more stringent assumptions. Both conditions could be satisfied if there is a priori knowledge about the missing data problem, such as knowing a specific subset of the variables should be used in the missing model. Alternatively, methods with the oracle property such as logistic regression with a SCAD or MCP penalty could be used to achieve such a rate. To use the nonparametric approach would require a priori knowledge or careful screening of the variables used in the model. In the simulations presented in Section 4 a sure independence screening method was used and results demonstrated that the weighted methods helped alleviate bias caused by missing data in the high-dimensional case.  

\par
\bigskip

\section{Variable Selection}

\subsection{Penalized Objective Function}

In this section the true linear covariates are not assumed to be known a priori. The following weighted and penalized objective function is used to rigorously estimate some of the linear coefficients as zero while accounting for the missing data, 
\begin{equation}
\label{penObjMissing}
\sumi \frac{R_i}{\hat{\pi}_i} \rho_\tau(Y_i - \vx_i^\top\vbeta - \vb(\vz_i)^\top\vxi) + \sum_{j=1}^p p_\lambda(|\beta_j|).
\end{equation}
The form of $\hat{\pi}_i$ depends on if parametric or nonparametric weights are used. Penalized objective functions are a popular alternative to best subset model selection methods such as BIC. One advantage is that penalized methods can be more computationally efficient than best subset methods, particularly when considering a large number of covariates. Tibshirani (1996) proposed the popular $L_1$ penalty (LASSO), $p_\lambda(|\beta|) = \lambda|\beta|$. However, to achieve model selection consistency with the LASSO penalty requires strong assumptions about the relationship between the active and inactive variables.(Zhao and Yu (2006)) Fan and Li (2001) proposed the SCAD penalty, motivated by finding a penalty function that provides an estimator with the oracle property. Zhang (2010) proposed MCP, another nonconvex penalty that has the oracle property. 

The SCAD penalty has the following form 
\begin{eqnarray*}
p_\lambda(|\beta|) &=& \lambda|\beta|I(0 \leq |\beta| < \lambda) + \frac{a \lambda |\beta| - \left(\beta^2 + \lambda^2\right)/2}{a-1}I(\lambda \leq |\beta| \leq a\lambda) \\
&&+ \frac{ (a + 1)\lambda^2}{2}I(|\beta| > a \lambda), \mbox{ for some } a > 2,
\end{eqnarray*}
while for the MCP penalty function,
\begin{eqnarray*}
p_\lambda(|\beta|) = \lambda \left( |\beta| - \frac{\beta^2}{2a\lambda} \right) I(0 \leq |\beta| < a \lambda) + \frac{a\lambda^2}{2}I\left(|\beta| \geq a \lambda\right), \mbox{ for some } a > 1.
\end{eqnarray*}

\begin{figure}
	\centering
		\includegraphics[width=.5\textwidth]{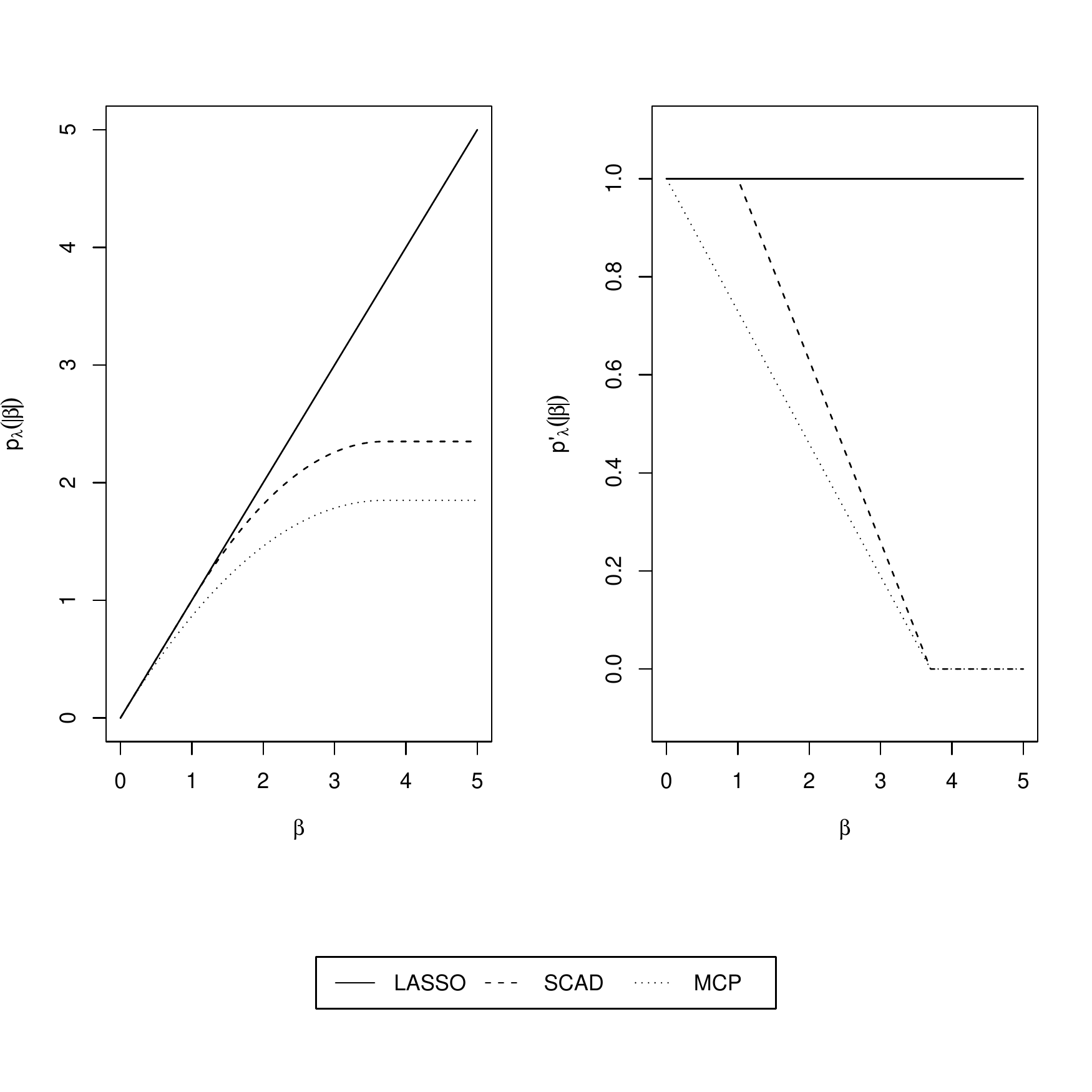}
	\caption{Plots of LASSO, SCAD, and MCP for $\lambda=1$ and $a=3.7$. On the left is the function $p_\lambda(|\beta|)$ and on the right is the plot of the derivatives.}
	\label{fig:scad}
\end{figure}

Figure \ref{fig:scad} plots the LASSO, SCAD and MCP functions and derivatives for $\lambda=1$ and $a=3.7$. The appeal of the nonconvex SCAD and MCP penalties is they do not over penalize larger coefficients with the derivatives going to zero as $|\beta|$ increases. A consequence of this property is the penalty function is not convex and therefore minimizing (\ref{penObjMissing}) is not a convex minimization problem and a local minimum is not guaranteed to be a global minimum. For both penalty functions, the tuning parameter $\lambda$ controls the complexity of the selected model and goes to zero as $n$ increases to $\infty$.

The oracle estimator with parametric weights is defined as $\tilde{\vbeta}^P \equiv \left(\mathbf{\hat{\vbeta}^{P^\top}}, \mathbf{0}_{p_n-q_n}^\top\right)^\top$ where
\begin{equation}
\label{weightedOracle}
\left(\mathbf{\hat{\vbeta}^{P}},\hat{\vxi}^{P}\right) = \underset{\left(\mathbf{\vbeta_1},\vxi\right)}{\argmin} \frac{1}{n} \sumi \frac{R_i}{\pi_i(\hat{\veta})} \rho_\tau\left(Y_i - \vx_i^\top\left(\mathbf{\vbeta_1}^\top,\mathbf{0}_{p-q}^\top\right)^\top - \vb(\vz_i)^\top\vxi\right).
\end{equation} 
Let $\tilde{\vbeta}^{K}$ be the similarly defined oracle estimator using the nonparametric weights. The oracle estimator sets to zero the coefficients for any linear covariates that do not have a relationship with the response. This estimators are the same as the estimators from (\ref{weighted_p}) and (\ref{weighted_n}) which only includes the active variables and therefore the asymptotic properties of the oracle estimators follow from Theorems \ref{xmar_part_lin_asy_normal_p}-\ref{o_high_d_n}. The high-dimensional setting requires an additional condition to control how quickly a nonzero signal can decay, this is not a concern when the dimension is fixed.

\begin{condition}
\label{cond_small_sig}
(Condition on the signal)
There exist positive constants $C_4$ and 
$C_5$ such that $2C_3 < C_4 < 1$ and $n^{(1-C_4)/2} \underset{1 \leq j \leq q_n}{\min} |\beta_{0j}| \geq C_5$.
\end{condition}  

The next two theorems state that asymptotically the oracle estimators are equivalent to a local minimum estimator of (\ref{penObjMissing}) using the MCP or SCAD penalty function.

\begin{theorem}
\label{thm_weighted_scad_p}
Assume Conditions \ref{cond_f}, \ref{cond_g_h}-\ref{xmar_cond_eta}, \ref{highd_cond_x}-\ref{cond_small_sig} are satisfied. Let $\mathcal{E}^P_n(\lambda)$ be the set of local minima of the the penalized objective function from (\ref{penObjMissing}) using parametric weights and the MCP or SCAD penalty function with tuning parameter $\lambda$. Let $\hat{\vOmega}^P \equiv \left(\tilde{\vbeta}^{P},\hat{\vxi}^P\right)$ be the oracle estimator for parametric weights. If 
$||\hat{\veta}-\veta_0|| = {\rm O}_{\rm p}\left(n^{-1/2}\right)$, $\lambda = {\rm o}\left(n^{-(1-C_4)/2}\right), n^{-1/2}q_n = o(\lambda)$, $n^{-1/2}k_n = {\rm o}(\lambda)$ and $\ln(p_n) = {\rm o}(n\lambda^2)$ as $n \rightarrow \infty$, then
\begin{equation*}
\Pr\left(\hat{\vOmega}^P \in \mathcal{E}^P_n(\lambda)\right) \rightarrow 1.
\end{equation*}
\end{theorem} 

\begin{theorem}
\label{thm_weighted_scad_n}
Assume Conditions \ref{cond_f}, \ref{cond_g_h}-\ref{xmar_cond_alpha}, \ref{xmar_cond_np}-\ref{cond_small_sig} are satisfied. Let $\mathcal{E}^K_n(\lambda)$ be the set of local minima of the the penalized objective function from (\ref{penObjMissing}) using nonparametric weights and the MCP or SCAD penalty function with tuning parameter $\lambda$. Let $\hat{\vOmega}^K \equiv \left(\tilde{\vbeta}^K,\hat{\vxi}^K\right)$ be the oracle estimator for nonparametric weights.  Assume the kernel based method has pointwise convergence of $\left| \tilde{\pi}_i-\pi_{i0} \right| = {\rm O}_{\rm p}\left(h^b+(h^sn)^{-1/2}\right)$ and a uniform convergence rate of $\underset{i}{\max} \left| \tilde{\pi}_i-\pi_{i0} \right| = {\rm O}_{\rm p}\left(h^b+(\ln n/h^sn)^{1/2}\right)$. If 
$\lambda = {\rm o}\left(n^{-(1-C_4)/2}\right)$, $n^{-1/2}q_n = {\rm o}(\lambda)$, $n^{-1/2}k_n = {\rm o}(\lambda)$, $h^b+(\ln n /(h^sn))^{1/2} = {\rm o}(\lambda)$ and $\ln(p_n) = {\rm o}(n\lambda^2)$ as $n \rightarrow \infty$, then
\begin{equation*}
\Pr\left(\hat{\vOmega}^K \in \mathcal{E}^K_n(\lambda)\right) \rightarrow 1.
\end{equation*}
\end{theorem} 

Theorem \ref{thm_weighted_scad_n} requires an additional assumption of $h^b+(\ln n /(h^sn))^{1/2} = o(\lambda)$ because of the uniform convergence rate of the kernel based estimator. Let $h = O\left(n^{-1/(b+s)}\right)$ and $\lambda = n^{-1/2+\delta}$ then conditions of Theorem \ref{thm_weighted_scad_n} hold if $\delta \in (\max(s/(2(b+s)),1/(2r+1), C_3),C_4/2)$, note $C_4/2>C_3$ by Condition \ref{cond_small_sig} and $C_4>1/2$, which is allowed by Condition \ref{cond_small_sig}, will gurantee that $C_4/2>1/(2r+1)$ as $r > 1.5$ by Condition \ref{cond_g_h} and that $C_4 > 2/(2(b+s))$ by Condition \ref{xmar_cond_np}. Then the conditions are satisfied if $p_n = \mbox{exp}(n^{\delta})$, thus allowing for exponential growth of the potential linear variables.  

The results are about local minimizers because the objective function is nonconvex. 
 These results hold for fixed dimensions under the conditions stated in Theorem \ref{xmar_part_lin_asy_normal_p} and \ref{xmar_part_lin_asy_normal_n}. Rates for $\lambda$ can be simplified to $n^{-1/2}k_n = o(\lambda)$ and $h^b+(\ln n /(h^sn))^{1/2} = o(\lambda)$, with the last rate only required when using nonparametric weights. The penalized objective function, using either SCAD or MCP, can be represented as a difference of convex functions. Tao and An (1997) present sufficient conditions for local optimization of the difference of convex functions. The strategy used in the proofs of Theorem \ref{thm_weighted_scad_p} and \ref{thm_weighted_scad_n} is to show that asymptotically the oracle estimator satisfies the sufficient conditions presented by Tao and An (1997) to be a local minimizer of (\ref{penObjMissing}). Proofs are provided in the appendix and in the supplementary material.

\

\section{Simulation studies}

\subsection{Solving the Penalized Estimator}
The weighted and penalized estimators are computed using the local linear approximation (LLA) algorithm (Zou and Li (2008)). Instead of directly solving (\ref{penObjMissing}) it is approximated with a sequence of convex objective functions
\begin{equation}
\label{q_scad_step}
\left( \hat{\vbeta}^t, \hat{\vxi}^t \right) = \underset{(\vbeta, \vxi)}{\argmin} \Big\{n^{-1}\sum_{i=1}^n \frac{R_i}{\hat{\pi}_i}\rho_\tau(Y_i - \vx_i^\top\vbeta - \vb(\vz_i)^\top\vxi)+ \sum_{j=1}^{p_n} p'_\lambda\left(|\hat{\vbeta}_j^{t-1}|\right)|\vbeta_j|\Big\},
\end{equation}
with the iteration ending once estimates $(\hat{\vbeta}^{t-1},\hat{\vxi}^{t-1})$ and $(\hat{\vbeta}^t, \hat{\vxi}^t)$ are sufficiently close, the simulations used $||\hat{\vbeta}^t - \hat{\vbeta}^{t-1}||_1 < 10^{-7}$. The first estimates, $t=1$, are obtained by setting $\hat{\vbeta}^0=0$, which is equivalent to starting with the LASSO penalty. The partial linear algorithm provided by Sherwood and Wang (2016) is adapted to handle the weighted objective function. The key observation is that $|u| = \rho_\tau(u) + \rho_\tau(-u)$ which allows (\ref{q_scad_step}) to be framed as a minimization problem with the objective function and penalty function using the same loss function of $\rho_\tau(\cdot)$. The CRAN package ``rqPen'' includes an implementation of this algorithm. (Sherwood and Maidman (2016))   

\par
\bigskip
\subsection{Simulation Setting and Results}

Monte Carlo simulations were performed to evaluate the finite sample size performance of the proposed variable selection method. 
The covariates are generated by
$\tilde{X} \sim N_{p-1}(0,\Sigma)$ where $\Sigma_{ij} = .7^{|i-j|}$, $X_u \sim U[0,\sqrt{12}]$, $X = [\tilde{X} \, X_u] \in \mathbb{R}^{n \times p}$, $Z_1 \sim U[0,1]$ and $Z_2 \sim U[-1,1]$. An upper bound of $\sqrt{12}$ is used for the linear uniform variable so that it has the same variance as the standard normal variables. Define $g_1(z) = \sin(2 \pi z)$ and $g_2(z) = z^3$. The data generating mechanism is
\begin{equation*}
Y_i =  x_{i1} - x_{i3} + x_{iu} + g_1(z_{i1}) + g_2(z_{i2}) + \epsilon_i.
\end{equation*}
In this section results are reported for the setting where $\epsilon_i \sim T_3$, for $\tau=.5$ and values of $p=8$ and 300. Supplementary material includes results for $\tau=.7$ when $\epsilon_i \sim (1+x_{iu})\vxi_i$ and $\vxi_i \sim \mathcal{N}(0,1)$. In the simulations $x_1$, $x_7$ and $z_2$ may have missing values. Two different missing models are considered
\begin{enumerate}
\item $\mbox{logit}(\Pr(R_i) = 1) = 1 + 2Y_i - 5x_{i2} + 5x_{i4} - 2z_{i1}$,
\item $\mbox{logit}(\Pr(R_i) = 1) = -2 + Y_i^3 + x_{i3}^2$.
\end{enumerate}
In each setting models are fit using parametric and nonparametric weights. There are only 3 variables with missing data and the remaining variables could be used to model $\Pr(R_i=1)$, while only a small number of variables actually influence the missing behavior. A glm sure independence screen is used to select the variables that are used in the missing model. (Fan and Song, 2010) A binomial glm is fit for each fully observed variable then ariables that are significant after a Bonferroni correction at the .05 level using a $\chi^2$ test are included in the final model. When screening for the nonparametric model the fully observed variable is transformed using a cubic B-spline with 2 internal knots.    

In all simulations a weight threshold of 25 was used, that is, any estimated weights larger than 25 are set to 25. This avoids the case of a very large weight being assigned to one observation which typically results in poor estimators and relates to Condition \ref{xmar_cond_alpha} which assumes that there is a lower bound to the probability that a subject would have complete data. The simulations focus on the benefits of using the weighted approach with the SCAD penalty. Results are reported for the estimator if all data were available (SCAD Full), an estimator that does not use weights (SCAD Naive), an estimator using parametric weights (SCAD P Wt) and an estimator using nonparametric weights (SCAD K Wt). Supplementary material included results for weighted versions of the LASSO and MCP penalty. As expected, the LASSO model tends to pick a larger model and SCAD and MCP have similar performance.  

Three hundred simulations were performed for each simulation setting. Tables 4.1-4.4 include the following statistics to summarize the results, where $m$ indexes the $m$th simulation and $p$ is the number of linear covariates considered. 
\begin{itemize}
\item  $\mathbf{r_n}$: Average number of observations with complete data. 
\item {\bf TV}: Average number of linear covariates correctly included in the model.
\item {\bf FV}: Average number of linear covariates incorrectly included in the model.
\item {\bf True}: Proportion of times the true model is exactly identified.
\item {\bf Bias}: $\sum_{j=1}^p \left|300^{-1} \sum_{m=1}^{300}\widehat{\vbeta}_j^m-\vbeta_{j0}\right|$.
\item {\bf MSE}:  $300^{-1} \sum_{j=1}^p \sum_{m=1}^{300}\left(\widehat{\vbeta}_j^m -\vbeta_{j0}\right)^2$.
\item {\bf AADE (Average Absolute Deviation Error)}: $300^{-1} \sum_{m=1}^{300} n^{-1} \sum_{i=1}^n \left| \hat{g}^m(\vz_i) - g_0(\vz_i)\right|$.
\end{itemize}

Zero to two internal knots for a cubic B-spline are considered for both nonlinear variables. Let $\nu = \nu_1 + k_{1n}+k_{2n}+6$ where $k_{1n}$ and $k_{2n}$ are the number of internal knots and 6 comes from the fact that 2 cubic basis splines are used and $\nu_1$ is the number of parametric terms included in the model. Then $\nu$ is the total number of parameters that are estimated for a given model. Let $\hat{\vbeta}_\lambda(k_{1n},k_{2n})$ and $\hat{\vxi}_\lambda(k_{1n},k_{2n})$ be the fits for a given $\lambda$, $k_{1n}$ and $k_{2n}$. The combintion of $\lambda$, $k_{1n}$ and $k_{2n}$ is selected by minimizing   

\begin{eqnarray*}
QBIC^W(\lambda, k_{1n}, k_{2n}) &=& \ln\left(\sum_{i=1}^n \frac{R_i}{\hat{\pi}_i} \rho_\tau(Y_i - \vx_i^\top\hat{\vbeta}_\lambda(k_{1n},k_{2n}) - \vb(\vz_i)^\top\hat{\vxi}_\lambda(k_{1n},k_{2n}))\right) \\
&+& \frac{\nu \ln(n)}{2n}.
\end{eqnarray*} 
For ``SCAD Naive" the weights of $R_i/\hat{\pi}_i$ are replaced with $R_i$, while for ``SCAD Full" a full data version is used without any weights. Horowitz and Lee (2005) proposed a similar BIC type method for a fully nonparametric additive quantile regression model. All of the SCAD based methods use $a=3.7$ as suggested by Fan and Li (2001).

Simulations were run with sample sizes of $200$, $400$ and $1000$. Kernel based estimates used Gaussian kernels and were estimated using the function \emph{kernesti.regr} from R pacakge ``regpro''. (Klemla (2013)) Bandwidth selection is done by using a suggestion made by Chen, Wan and Zhou (2015) for a simplified approach of Sepanski et al. (1994). The bandwidth $h=\hat{\sigma}_{\vt}n^{-1/(s+2)}$, where $\hat{\sigma}_{\vt}$ is the estimated standard deviation of the covariates used in the missing model and 2 comes from the order of the Gaussian kernel. Tables 4.1 and 4.2 report results for p=8 using missing models 1 and 2 respectively. While Table 4.3 and 4.4 report the results for p=300. Results in Tables 4.1 and 4.2 demonstrate that there is clear bias for the small p case, $p=8$. Both weighted methods reduce the bias and for sufficiently large sample sizes provide smaller MSE and AADE. Table 4.1 presents results when the parametric relationship for weights holds. In keeping with Remark 1 after Theorems 1 and 2 both the parametric weights and nonparametric weights perform about the same even if the parametric model holds. When the fitted parametric model is misspecified, as in the simulations corresponding to Table 4.2, there is a clear advantage to using the kernel based estimates. However, when considering results with heteroscedastic errors as presented in the supplementary material there are settings where the parametric model will perform better when the parametric missing model holds. Suggesting that there are instances where the finite sample performance of the parametric method would be superior to the kernel based method. 

Tables 4.3 and 4.4 present the case for $p=300$ for the parametric and nonparametric missing models. In Table 4.4, the case where the parametric model is misspecified, the kernel weights provide a clear advantage. In the high dimensional cases the naive method does well for model selection, but the estimated coefficients are biased. The weighted methods reduce the bias and increase accuracy for the linear and nonlinear parts of the model. 

\begin{table}[ht]
\centering
\begin{tabular}{lllllllll}
  \hline
Method & n & $\mathbf{r_n}$ & TV(3) & FV & True & Bias & MSE & AADE \\ 
  \hline
SCAD Full   & 200  & 200  & 2.87 & 0.01 & 0.92 & 0.16 & 0.20 & 0.30 \\ 
  SCAD Full & 400  & 400  & 2.98 & 0.00 & 0.99 & 0.04 & 0.04 & 0.20 \\ 
  SCAD Full & 1000 & 1000 & 3.00 & 0.00 & 1.00 & 0.01 & 0.01 & 0.13 \\ 
   \hline
SCAD Naive   & 200  & 140 & 2.67 & 0.24 & 0.69 & 0.69 & 0.56 & 0.56 \\ 
  SCAD Naive & 400  & 281 & 2.84 & 0.15 & 0.83 & 0.44 & 0.30 & 0.46 \\ 
  SCAD Naive & 1000 & 705 & 2.96 & 0.06 & 0.94 & 0.21 & 0.09 & 0.38 \\ 
   \hline
SCAD P Wt   & 200  & 140 & 2.80 & 0.36 & 0.64 & 0.55 & 0.48 & 0.50 \\ 
  SCAD P Wt & 400  & 281 & 2.95 & 0.12 & 0.87 & 0.22 & 0.16 & 0.39 \\ 
  SCAD P Wt & 1000 & 705 & 3.00 & 0.02 & 0.98 & 0.09 & 0.03 & 0.27 \\ 
   \hline
SCAD K Wt   & 200  & 140 & 2.87 & 0.26 & 0.75 & 0.43 & 0.32 & 0.45 \\ 
  SCAD K Wt & 400  & 281 & 2.94 & 0.12 & 0.88 & 0.27 & 0.16 & 0.36 \\ 
  SCAD K Wt & 1000 & 705 & 2.99 & 0.02 & 0.98 & 0.12 & 0.03 & 0.28 \\ 
   \hline
\end{tabular}
\caption{Results for $\epsilon \sim T_3$ with p=8 and $\tau = .5$ when parametric missing model holds (missing data model 1)} 
\end{table}

\begin{table}[ht]
\centering
\begin{tabular}{lllllllll}
  \hline
Method & n & $\mathbf{r_n}$ & TV(3) & FV & True & Bias & MSE & AADE \\ 
  \hline
SCAD Full   & 200  & 200  & 2.84 & 0.02 & 0.91 & 0.18 & 0.21 & 0.31 \\ 
  SCAD Full & 400  & 400  & 2.99 & 0.00 & 0.99 & 0.01 & 0.03 & 0.21 \\ 
  SCAD Full & 1000 & 1000 & 3.00 & 0.00 & 1.00 & 0.01 & 0.01 & 0.13 \\ 
   \hline
SCAD Naive & 200    & 142 & 2.73 & 0.04 & 0.83 & 0.50 & 0.37 & 0.41 \\ 
  SCAD Naive & 400  & 284 & 2.95 & 0.00 & 0.96 & 0.27 & 0.13 & 0.27 \\ 
  SCAD Naive & 1000 & 706 & 3.00 & 0.00 & 1.00 & 0.21 & 0.02 & 0.20 \\ 
   \hline
SCAD P Wt   & 200  & 142 & 2.97 & 0.05 & 0.94 & 0.25 & 0.11 & 0.38 \\ 
  SCAD P Wt & 400  & 284 & 3.00 & 0.00 & 1.00 & 0.19 & 0.03 & 0.27 \\ 
  SCAD P Wt & 1000 & 706 & 3.00 & 0.00 & 1.00 & 0.21 & 0.02 & 0.20 \\ 
   \hline
SCAD K Wt   & 200  & 142 & 2.96 & 0.06 & 0.92 & 0.22 & 0.12 & 0.39 \\ 
  SCAD K Wt & 400  & 284 & 3.00 & 0.00 & 1.00 & 0.09 & 0.03 & 0.25 \\ 
  SCAD K Wt & 1000 & 706 & 3.00 & 0.00 & 1.00 & 0.11 & 0.01 & 0.16 \\ 
   \hline
\end{tabular}
\caption{Results for $\epsilon \sim T_3$ with $p=8$ and $\tau = .5$ when nonparametric missing model holds (missing data model 2)} 
\end{table}

\begin{table}[ht]
\centering
\begin{tabular}{lllllllll}
  \hline
Method & n & $\mathbf{r_n}$ & TV(3) & FV & True & Bias & MSE & AADE \\ 
  \hline
SCAD Full   & 200  & 200  & 2.81 & 0.32 & 0.68 & 0.25 & 0.26 & 0.30 \\ 
  SCAD Full & 400  & 400  & 2.96 & 0.00 & 0.97 & 0.08 & 0.09 & 0.21 \\ 
  SCAD Full & 1000 & 1000 & 3.00 & 0.00 & 1.00 & 0.01 & 0.01 & 0.13 \\ 
   \hline
SCAD Naive & 200    & 141 & 2.61 & 1.50 & 0.30 & 0.80 & 0.65 & 0.56 \\ 
  SCAD Naive & 400  & 281 & 2.80 & 0.22 & 0.76 & 0.49 & 0.35 & 0.46 \\ 
  SCAD Naive & 1000 & 704 & 2.95 & 0.06 & 0.94 & 0.23 & 0.12 & 0.39 \\ 
   \hline
SCAD P Wt & 200    & 141 & 2.79 & 6.45 & 0.10 & 0.81 & 0.64 & 0.46 \\ 
  SCAD P Wt & 400  & 281 & 2.89 & 2.63 & 0.38 & 0.39 & 0.27 & 0.38 \\ 
  SCAD P Wt & 1000 & 704 & 2.99 & 0.50 & 0.77 & 0.11 & 0.04 & 0.27 \\ 
   \hline
SCAD K Wt & 200    & 141 & 2.79 & 3.92 & 0.19 & 0.68 & 0.51 & 0.45 \\ 
  SCAD K Wt & 400  & 281 & 2.93 & 0.51 & 0.69 & 0.29 & 0.19 & 0.34 \\ 
  SCAD K Wt & 1000 & 704 & 2.99 & 0.02 & 0.98 & 0.12 & 0.03 & 0.29 \\ 
   \hline
\end{tabular}
\caption{Results for $\epsilon \sim T_3$ with $p=300$ and $\tau=.5$ when parametric missing model holds (missing data model 1)} 
\end{table}

\begin{table}[ht]
\centering
\begin{tabular}{lllllllll}
  \hline
Method & n & $\mathbf{r_n}$ & TV(3) & FV & True & Bias & MSE & AADE \\ 
  \hline
SCAD Full & 200    & 200  & 2.86 & 0.34 & 0.69 & 0.16 & 0.21 & 0.30 \\ 
  SCAD Full & 400  & 400  & 2.99 & 0.00 & 0.99 & 0.02 & 0.04 & 0.21 \\ 
  SCAD Full & 1000 & 1000 & 3.00 & 0.00 & 1.00 & 0.00 & 0.01 & 0.13 \\ 
   \hline
SCAD Naive   & 200  & 141 & 2.68 & 1.30 & 0.32 & 0.61 & 0.44 & 0.41 \\ 
  SCAD Naive & 400  & 282 & 2.92 & 0.06 & 0.90 & 0.29 & 0.13 & 0.29 \\ 
  SCAD Naive & 1000 & 707 & 3.00 & 0.00 & 1.00 & 0.22 & 0.04 & 0.20 \\ 
   \hline
SCAD P Wt   & 200  & 141 & 2.91 & 3.10 & 0.25 & 0.43 & 0.22 & 0.40 \\ 
  SCAD P Wt & 400  & 282 & 3.00 & 0.21 & 0.86 & 0.21 & 0.04 & 0.28 \\ 
  SCAD P Wt & 1000 & 707 & 3.00 & 0.00 & 1.00 & 0.20 & 0.02 & 0.20 \\ 
   \hline
SCAD K Wt   & 200  & 141 & 2.91 & 3.90 & 0.18 & 0.43 & 0.23 & 0.39 \\ 
  SCAD K Wt & 400  & 282 & 3.00 & 0.36 & 0.80 & 0.13 & 0.03 & 0.26 \\ 
  SCAD K Wt & 1000 & 707 & 3.00 & 0.00 & 1.00 & 0.09 & 0.01 & 0.16 \\ 
   \hline
\end{tabular}
\caption{Results for $\epsilon \sim T_3$ with $p=300$ and $\tau=.5$ when parametric missing model holds (missing data model 2)} 
\end{table}

\par

\

\section{Data Analysis}
The proposed methods are applied to model time sober for patients leaving a rehabilitation center. The data is from the UIS study described in Section 1.3 of Hosmer et al. (2008). The covariates consist of binary and quantitative variables. The binary variables are: non-white race, treatment site (A or B), cocaine use and randomized treatment assignment (long or short). The quantitative variables are:  age, Beck Depression score at admission, number of prior drug treatments and length of stay. The sample size is 628 with 53 samples having missing data. Age, Beck score, number of prior treatments, non-white race, cocaine use, heroin use and IV use all have missing values. Randomized treatment, treatment site, length of stay and time sober are fully observed. Table \ref{tabLog} summarizes a logistic regression with the missing indicator as the response and the fully observed variables as the predictors. Patients that stay longer are less likely to have missing data, which shows the missing completely at random assumption does not hold.

\begin{table}[ht]
\centering
\begin{tabular}{rrrrr}
  \hline
 & Estimate & Std. Error & z value & $\Pr$($>$$|$z$|$) \\
  \hline
(Intercept) & 1.02 & 0.284 & 3.61 & 0.0003 \\
  Randomized Treatment (Long)    & 0.12 & 0.301 & 0.40 & 0.6893 \\
  Treatment Site (B)     & 0.50 & 0.388 & 1.30 & 0.1943 \\
  Length of Stay       & 0.02 & 0.004 & 4.74 & 0.0000 \\
  Time Sober & 0.00 & 0.001 & 0.06 & 0.9526 \\
   \hline
\end{tabular}
\caption{Logistic Regression Model of missingness}
\label{tabLog}
\end{table}  

A weighted quantile regression extension of BIC (Schwarz (1978)) was used to determine if a non-binary covariate would be modeled as a linear or nonlinear covariate. For a model $\nu_\tau$ of the $\tau$th quantile define the weighted quantile regression BIC (WQBIC) as
\begin{equation}
\label{q_screen}
\mbox{WQBIC}(\nu_\tau) = \ln\left(\sum_{i=1}^{n} \frac{R_i}{\hat{\pi}} \rho_\tau(Y_i-\hat{Y}_{i}(\nu_\tau))\right) + \tilde{p}(\nu_\tau)\frac{\ln(n)}{2n},
\end{equation}
where $\hat{Y}_{i}(\nu_\tau)$ represents the fitted value of the $\tau$th quantile for the $i$th sample and $\tilde{p}(\nu_\tau)$ is the number of coefficients estimated in model $\nu_\tau$. To determine if a covariate will be included as a linear or nonlinear variable seven models are considered, an intercept only model and six models with the covariate as the only predictor. The six univariate predictor models are a linear model and 5 nonlinear models, using B-splines with 0,1,2,3 or 4 internal knots. An intercept only model is included to protect against a nonlinear model being selected by random chance. If the linear or intercept model minimizes (\ref{q_screen}) then the covariate is included as a linear predictor. The intercept only model indicates that the variable does not have a strong signal, but is included for variable selection as it may be conditionally important. Otherwise the variable is fit as a nonlinear variable using the number of internal knots that corresponds with the model that minimizes (\ref{q_screen}). Variables were assigned as linear or nonlinear using WQBIC for the .05 and .95 quantiles using a naive approach and parametric and nonparametric weights. None of the variables are selected as nonlinear at the .05 quantile, while at the .95 quantile Beck score and length of stay are selected as nonlinear variables. Thus suggesting that there are reasons to model each quantile separately and not impose a global model. The designation of linear and nonlinear variables is the same for all three methods.

To assess the proposed weighted objective function (\ref{penObjMissing}) the data is randomly partitioned into a testing data set with 100 observations and a training data set with 528 observations, with models fit for the .05 and .95 quantiles using the training data. First, the nonlinear variables are designated using the WQBIC approach outline in the previous paragraph. Then the data is fit using the SCAD penalized objective function as outlined in the simulations section. In each training set we fit models using a naive approach and parametric and nonparametric weights. The .05 and .95 conditional quantile models are then used to estimate a 90\% prediction interval for the training data set. The process was repeated 500 times, each with a new random partition of the data. The capture rate, percentage of time the prediction interval captured the true value, and average prediction interval lengths are reported in Table \ref{tabDataResults} with the standard deviation of interval length reported in parentheses. The weighted methods have capture rates of .89, while the naive method has a capture rate of .88. All methods are close to the  expected coverage of .90 and have almost identical average interval lengths. Similar results were found when estimating a 60\% or 95\% confidence intervals and those results are available in the supplementary material. Even after the variability from model selection both methods are providing consistent prediction intervals. However, the proposed weighted methods are not dramatically outperforming the naive approach. This is because the rate of missingness is low, around 8.5\%, and the missingness does not depend on the response. For the last point one can see in Table \ref{tabLog} that Time Sober was the least influential variable when modeling missingness. Even if the weighted method does not provide a drastically different solution than the naive method it is a useful tool to check if ignoring the missing data structure is resulting in a biased analysis.

\begin{table}[ht]
\centering
\begin{tabular}{lrr}
  \hline
Method & Capture Rate & Interval Length \\ 
  \hline
Naive & 0.88 & 468.47(89.23) \\ 
  Parametric Weights & 0.89 & 468.04(89.98) \\ 
  Kernel Weights & 0.89 & 469.59(89.28) \\ 
   \hline
\end{tabular}
\caption{Random partition 90\% prediction interval results}
\label{tabDataResults}
\end{table}

\

\section{Discussion}

This paper investigates variable selection for partial linear quantile regression models with missing covariates. B-splines are used for estimating nonlinear relationships with the response, IPW is used to handle bias caused by missing covariates and a nonconvex penalty is used to perform simultaneous estimation and variable selection. Possible extensions will be discussed here. 

The theory presented considers the internal knots to be fixed values. This ensures that the value of $\vb(\vz_i)$ does not change depending on which variables have been observed. However, in practice sample quantiles, or other data driven methods, are typically used to select the internal knots. Thus the B-spline used will vary depending on if all of the data was observed or not. In the simulations and applied analysis internal knots were estimated sample quantiles of the observed values. The simulations showed that the weighted approach improves the accuracy of the nonlinear fits. How to optimally select the internal knots in the presence of missing data is an interesting question that merits future research. 

A challenging problem, particularly for high dimensional data, is the problem of identifying which covariates are nonlinear or linear terms. Solutions for handling this problem for semiparametric mean regression have been proposed by Zhang, Cheng and Liu (2011); Huang, Wei and Ma (2012); Lian, Liang and Ruppert (2015). Linear and nonlinear terms can be included in the model by including the original variable as a linear term and a basis spline of the original variable as the nonlinear term. Then model selection of the linear and nonlinear terms can be done by using a penalty for the linear terms and a group penalty for the nonlinear terms. Extending this approach to quantile regression is an area of research I plan on covering in depth in the future. 

\


\newcommand{\norm}[1]{\left\lVert #1 \right\rVert}

\section{Appendix}
\noindent {\large \bf Proofs} \\
\noindent{\bf Definition and properties of theoretically centered basis functions} \\
Throughout the appendix $C$ is used to denote a positive constant which does not depend on $n$ and may vary from 
line to line. For a vector $\vx$, $\norm{\vx}$ denotes its Euclidean norm and
for a matrix $A$, $||A||= \sqrt{\lambda_{\max}(A'A)}$ denotes its spectral norm.

Proofs for the theorems using weights derived from kernel smoothing estimates are provided because these proofs are more challenging and the parametric proofs use similar techniques. For the sake of simplicity it is assumed that the kernel function, $K(\cdot)$, is symmetric, that $s=1$ and $b=2$. Assuming the kernel is symmetric and $b=2$ means that the kernel is a symmetric pdf which is a common class of kernel functions. The assumption of $s=1$ is to simplify the Taylor expansions used in results about kernel functions. In addition the proofs typically focus on the more general case of where $q_n$ and $p_n$ can increase with $n$. One exception is there is a proof for asymptotic normality from Theorem \ref{xmar_part_lin_asy_normal_n} because the paper does not contain an exact high-dimensional generalization of this result. Throughout the appendix $C$ is used to represent a generic positive constant that can change from line to line. 

For ease of proof the B-spline basis functions are theoretically centered similar to the approach used in Xue and Yang (2006). Specifically, $b_j(\cdot)$ from Section 2.1 is transformed by defining
\begin{equation*}
w_j(z_{ik}) = \sqrt{k_n}\left( b_{j}(z_{ik}) - \frac{{\rm E}\left\{b_{j}(z_{ik})\right\}}{{\rm E}\left\{b_0(z_{ik})\right\}}b_0(z_{ik})\right)
\end{equation*}
For a given covariate $z_{ik}$, let $\vw(z_{ik})= \left(w_1(z_{ik}),\ldots,w_{L_n}(z_{ik})\right)^\top$
be the vector of basis functions, and $\vWzi$ denote the  $J_n$-dimensional vector
$\left(1,\vw(z_{i1})^\top,\ldots,\vw(z_{id})^\top\right)^\top$, where $J_n=dL_n+1$ and $W = \left(\vW(\vz_1),\ldots,\vW(\vz_n)\right)^\top \in \mathbb{R}^{n \times J_n}$. Let 
\begin{equation*}
\left(\hat{\vbeta}^{{K}^*}_1, \hat{\vgamma}^K\right) = \underset{(\vbeta_1, \vgamma)}{\argmin} \sumi \frac{R_i}{\tilde{\pi}_i}\rho_\tau(Y_i - \vx_{A_i}^\top\vbeta_1 - \vWzi^\top\vgamma),
\end{equation*}
then $\hat{\vbeta}^{{K}^*}_1=\hat{\vbeta}_1^{K}$ and $\hat{g}^K(z_i) = \vWzi^\top\hat{\vgamma}^K$. Thus the alternative basis for the nonlinear terms does not alter the estimates of the linear coefficients or the unknown additive function. To help analyze the asymptotic behavior of $\hat{\vbeta}_1^K$, while accounting for the estimation of $\hat{\vgamma}^K$, following the techniques of He et al. (2002), define:
\begin{eqnarray*}
D_n &=& \mbox{diag}\left(f_i(0 \mid \vx_i,\vz_i)R_i\pi_{i0}^{-1}\right) \in \mathbb{R}^{n \times n}, \\
\tilde{D}_n &=& \mbox{diag}\left(f_i(0 \mid \vx_i \vz_i)R_i\tilde{\pi}_i^{-1}\right) \in \mathbb{R}^{n \times n}, \\
W&=& (\vW(\vz_1),\ldots,\vW(\vz_n))^\top \in \mathbb{R}^{n \times J_n},\\
P &=& W(W^\top\tilde{D}_nW)^{-1}W^\top\tilde{D}_n \in \mathbb{R}^{n \times n}, \\
X^* &=&(\vx_1^*,\ldots,\vx_n^*)^\top=(I_n-P)X_A \in \mathbb{R}^{n \times q_n},\\
W_D^2 &=& W^\top\tilde{D}_nW \in  \mathbb{R}^{J_n \times J_n} , \\
\vtheta_1 &=& \sqrt{n}\left(\vbeta_1-\vbeta_{10}\right)\in \mathbb{R}^{q_n}, \\
\vtheta_2 &=& W_D\left(\vgamma-\vgamma_0\right) + W_D^{-1}W^\top D_nX_A(\vbeta_1-\vbeta_{10}) \in \mathbb{R}^{J_n}, \\
\tilde{\vx}_i &=& n^{-1/2}\vx_i^* \in  \mathbb{R}^{q_n}, \\
\vtWzi &=& W_D^{-1}\vWzi \in \mathbb{R}^{J_n},\\
\tilde{\vs}_i &=& \left(\tilde{\vx}_i^\top,\vtWzi\right)^\top \in \mathbb{R}^{q_n+J_n},\\
u_{ni}&=&\vWzi^\top\vgamma_{0} - g_{0}(\vzi),\\
Q_i(a_n)  &=& \rho_\tau\left(\epsilon_i - a_n\tilde{\vx}_i^\top\vtheta_1 - a_n\tilde{\vW}(\vz_i)^\top\theta_2 - u_{ni}\right), \\
{\rm E}_s\{Q_i\} &=& {\rm E}\left\{Q_i \mid \vx_i,\vz_i\right\}.
\end{eqnarray*}
Notice that
\begin{equation*}
\sumi \frac{R_i}{\tilde{\pi}_i} \rho_\tau(Y_i - \vx_{A_i}^\top\vbeta_1 - \vWzi^\top\vgamma) = \sumi \frac{R_i}{\tilde{\pi}_i} \rho_\tau(\epsilon_i - \tilde{\vx}_i^\top\vtheta_1 - \vtWzi^\top\vtheta_2 - u_{ni}).
\end{equation*}
Defining $(\hat{\vtheta}_1,\hat{\vtheta}_2) = \underset{(\vtheta_1,\vtheta_2)}{\argmin} \sumi \left(R_i/\tilde{\pi}_i\right) \rho_\tau(\epsilon_i - \tilde{\vx}_i^\top\vtheta_1 - \vtWzi^\top\vtheta_2 - u_{ni})$ then $\hat{\vtheta}_1 = \sqrt{n}\left(\hat{\vbeta}_1^K-\vbeta_{10}\right)$
and $\hat{\vtheta}_2 = W_D\left(\hat{\vgamma}^K-\vgamma_0\right) + W_D^{-1}W^\top\tilde{D}_nX_A(\hat{\vbeta}_1^K-\vbeta_{10})$. Define $d_n = q_n + J_n$, to simplify the notation when examining estimation for the growing number of splines and the potentially growing number of covariates.     
\newline

Lemma 1 of Chen, X., Wan, A. and Zhou, Y. (2015) is restated, which provides a key result for working with the kernel based estimates. 

\begin{lemma}
Assume Condition \ref{xmar_cond_np} holds. Assume $\phi(\cdot)$ is a real valued function, $\vu_1$,\ldots,$\vu_n$ are independent variables different from $t_1$,\ldots,$t_n$. Define $f_{t_i,\vu_i}(t,\vu)$ as the joint density of $(t_i,\vu_i)$. For some $s$ such that $r < 1 - s^{-1}$ assume ${\rm E}\left|\phi(t_i,\vu_i)\right|^s < \infty$ and $\underset{t}{\mbox{sup}} \int |\phi(t_i,\vu_i)|^s f_{t_i,\vu_i}(t,\vu) d\vu < \infty$. Then as $nh^{2r-1} \rightarrow \infty$
\begin{eqnarray*}
&& \underset{t}{\mbox{sup}} \left| \frac{1}{nh} \sumi \left\{ K\left(\frac{t_i-t}{h}\right)K\left(\frac{t_i-t}{h}\right)^k \phi(t_i,\vu_i) - {\rm E}\left\{K\left(\frac{t_i-t}{h}\right)K\left(\frac{t_i-t}{h}\right)^k \phi(t_i,\vu_i)\right\} \right\} \right|\\
&=& {\rm O}_{\rm p}\left( \left(\frac{\ln h^{-1}}{nh} \right)^{1/2} \right).   
\end{eqnarray*}
\end{lemma} 
If ${\rm E}\left\{\phi(t_i,\vu_i)\right\}$ is continuous and twice differentiable at $t_i=t$ then for a symmetric kernel function 
\begin{equation*}
\underset{t}{\mbox{sup}} \left| \frac{1}{nh} \sumi K\left(\frac{t_i - t}{h}\right) \phi(t_i,\vu_i) 
- f(t)\frac{1}{n} \sumi {\rm E}\left\{\phi(t_i,\vu_i) \mid t_i = t \right\} \right| = {\rm O}_{\rm p}\left(h^2 + \left(\frac{\ln h^{-1}}{nh} \right)^{1/2} \right). 
\end{equation*}
\begin{proof}
See proof of Lemma 1 in supplementary material of Chen, X., Wan, A. and Zhou, Y. (2015).
\end{proof}

The next lemma establishes important rates for the spline basis. 
\begin{lemma}\label{spline}
The spline basis vector has the following properties.\\
(1) ${\rm E}\{||\vWzi||\}\leq b_1\sqrt{k_n}$, $\forall \ i$, for some positive constant $b_1$ for 
all $n$ sufficiently large.\\
(2) There exists some positive constants $b_2$ and $b_3$ such that for all $n$ sufficiently large
${\rm E}\{\lambda_{\min}(\vWzi\vWzi^T)\}\geq b_2$ and ${\rm E}\{\lambda_{\max}(\vWzi\vWzi^T)\}\leq b_3$.\\
(3) ${\rm E}\{||W_D^{-1}||\}\geq b_4n^{-1/2}$,
for some positive constant $b_4$, for all $n$ sufficiently large.\\
(4) 
$\underset{i}{\max} ||\vtWzi||={\rm O}_{\rm p}\left(\sqrt{k_n/n}\right)$.\\
\end{lemma}
\begin{proof}
These results follow from Lemma 1 of Sherwood and Wang (2016), that Condition \ref{xmar_cond_alpha} provides that $\pi_{i0}$ has a uniform lower and upper bound and that the uniform convergence rate of the kernel based estimates provides the estimated weights will have a lower and upper bound with probability approaching one.
\end{proof}

Proofs for the following lemmas are provided in the online supplementary material. 

\begin{lemma}
\label{x_star}
If Conditions 1-3 and 7-8 are satisfied,  then
$n^{-1/2}X^* = n^{-1/2}\Delta_n + {\rm o}_{\rm p}(1)$.  Furthermore, $n^{-1}{X^*}^\top\tilde{D}_nX^* = \Sigma_1 + {\rm o}_{\rm p}(1)$, where $\Sigma_1$ is defined in Condition 6.
\end{lemma}

\begin{lemma}
\label{qi_rate}
If the conditions of Theorem 4 hold then for any $\omega > 0$
\begin{equation*}
\Pr\left( \underset{||\vtheta||=L}{\inf} d_n^{-1} \sumi \frac{R_i}{\tilde{\pi}_i} (Q_i(\sqrt{d_n}) - Q_i(0)) > 0\right) \geq 1-\omega.
\end{equation*}
\end{lemma}

\noindent{\bf Proof of nonlinear convergence rate for Theorem \ref{o_high_d_n}}
\begin{proof}
It follows from Lemma \ref{qi_rate} that
$\left|\left|W_D(\hat{\vgamma}^K-\vgamma_0)\right|\right| = {\rm O}_{\rm p}\left(\sqrt{q_n+k_n}\right)$. From Schumaker (1981, p.227) it follows that $\underset{i}{\max} |u_{ni}|={\rm O}\left(k_n^{-r}\right)$. 
Combining this with Condition \ref{cond_g_h} then
\begin{eqnarray*}
n^{-1}\sumi f_i(0 \mid \vx_i, \vz_i)\left(\hat{g}(\vz_i)-g_0(\vz_i)\right)^2 &=& n^{-1} \sumi f_i(0 \mid \vx_i,\vz_i)\left(\vW(\vz_i)^\top(\hat{\gamma}-\gamma_0) - u_{ni}\right)^2 \\
&\leq& n^{-1} \left(\hat{\vgamma}-\vgamma_0\right)W_D^2\left(\hat{\vgamma}-\vgamma_0\right) + {\rm O}_{\rm p}\left(n^{-2r/(2r+1)}\right)\\
&=& {\rm O}_{\rm p}\left(\frac{q_n}{n} + n^{-2r/(2r+1)}\right). 
\end{eqnarray*}
Proof is complete by Condition 1 which provides a uniform lower and upper bound for $f_i(0 \mid \vx_i,\vz_i)$. 
\end{proof}

For $\hat{\vbeta}_1^K$ the asymptotic normality as stated in Theorem \ref{xmar_part_lin_asy_normal_n} is proved. The rate of convergence from Theorem \ref{o_high_d_n} is a generalization of that result for the case of growing $q_n$. 
\\
\noindent{\bf Proof of asymptotic normality from Theorem \ref{xmar_part_lin_asy_normal_n}}
\begin{proof}
It follows from Lemma A.4 of the supplementary material that $\sqrt{n}(\hat{\beta_1}^K - \beta_0)
= (n^{-1}\sumi f_i(0 \mid \vx_i,\vz_i) \vx_i^* {\vx_i^*}^\top)^{-1}n^{-1/2}\sumi (R_i/\tilde{\pi}_i)\vx_i^*\psi_\tau(\epsilon_i) + {\rm o}_{\rm p}(1).$

Using Lemma \ref{x_star} for the first equality and Lemma A5 from the supplementary material for the second
\begin{eqnarray*}
\sqrt{n}(\hat{\beta_1}^W - \beta_0) &=& \left(\Sigma_1+{\rm o}_{\rm p}(1)\right)^{-1} n^{-1/2} \sumi \frac{R_i}{\tilde{\pi}_i} \delta_i \psi_\tau(\epsilon_i)(1+{\rm o}_{\rm p}(1)) \\
&=& \left(\Sigma_1+{\rm o}_{\rm p}(1)\right)^{-1}\left[\frac{1}{\sqrt{n}} \sumi \frac{R_i}{\pi_{i0}} \vdelta_i \psi_\tau(\epsilon_i) - \frac{1}{\sqrt{n}} \sumi \frac{R_i-\pi_{i0}}{\pi_{i0}}{\rm E}\{\vdelta_i \psi_\tau(\epsilon_i) \mid \vt_i\}  \right](1+{\rm o}_{\rm p}(1)).
\end{eqnarray*}
The two sums have expected values of zero. To complete the proof check the variance of the two sums and their covariance. The variance of the first sum is 
\begin{eqnarray*}
{\rm Var}\left\{\frac{R_i}{\pi_{i0}} \vdelta_i \psi_\tau(\epsilon_i)\right\} &=&
{\rm E}\left\{\frac{R_i}{\pi_{i0}^2} \vdelta_i\vdelta_i^\top \psi_\tau(\epsilon_i)^2\right\} \\
&=& {\rm E}\left\{\frac{1}{\pi_{i0}} \vdelta_i\vdelta_i^\top \psi_\tau(\epsilon_i)^2\right\}. \\
\end{eqnarray*}
The variance of the second sum is 
\begin{eqnarray*}
{\rm Var}\left\{\frac{R_i-\pi_{i0}}{\pi_{i0}}{\rm E}\{\vdelta_i \psi_\tau(\epsilon_i) \mid \vt_i\}\right\}
&=& {\rm E}\left\{\frac{\left(R_i-\pi_{i0}\right)^2}{\pi_{i0}^2}{\rm E}\{\vdelta_i \psi_\tau(\epsilon_i) \mid \vt_i\} {\rm E}\{\vdelta_i^\top \psi_\tau(\epsilon_i) \mid \vt_i\}\right\} \\
&=& {\rm E}\left\{\frac{1-\pi_{i0}}{\pi_{i0}}{\rm E}\{\vdelta_i \psi_\tau(\epsilon_i) \mid \vt_i\} {\rm E}\{\vdelta_i^\top \psi_\tau(\epsilon_i) \mid \vt_i\}\right\}.
\end{eqnarray*}
For the covariance of the sums use the assumption that $\pi_{i0}$ is known given $\vt_i$ and the law of iterated expectations to get
\begin{eqnarray*}
{\rm Cov}\left\{\frac{R_i}{\pi_{i0}} \vdelta_i \psi_\tau(\epsilon_i),\frac{R_i-\pi_{i0}}{\pi_{i0}}{\rm E}\{\vdelta_i \psi_\tau(\epsilon_i) \mid \vt_i\}\right\} &=& {\rm E}\left\{\frac{R_i(R_i-\pi_{i0})}{\pi_{i0}^2} \vdelta_i \psi_\tau(\epsilon_i){\rm E}\{\vdelta_i^\top \psi_\tau(\epsilon_i) \mid \vt_i\} \right\} \\
&=& {\rm E}\left\{\frac{1-\pi_{i0}}{\pi_{i0}} {\rm E}\left\{ \vdelta_i \psi_\tau(\epsilon_i) \mid \vt_i\right]{\rm E}\{\vdelta_i^\top \psi_\tau(\epsilon_i) \mid \vt_i\} \right\}.
\end{eqnarray*}

\end{proof}

\noindent{\bf Proof of Theorem 6.}
\begin{proof}
Consider the unpenalized objective function for the oracle model
\begin{eqnarray*}
S_n(\vbeta_1,\vgamma) = \frac{1}{n} \sumi \frac{R_i}{\tilde{\pi}_i} \rho_\tau(Y_i-\vx_i^\top(\vbeta_1,\mathbf{0}_{p-q}) - \vW(\vz_i)^\top\vgamma).
\end{eqnarray*}
Define 
\begin{eqnarray*}
&& (\bar{w}_0(\vz_i),\bar{w}_1(\vz_i),\ldots,\bar{w}_{L_n}(\vz_i),\ldots,\bar{w}_{(d-1)L_n+1}(\vz_i),\ldots,\bar{w}_{dL_n}(\vz_i)) \\
&=& (1,w_1(z_{i1}),\ldots,w_{k_n+l}(z_{i1}),\ldots,w_{1}(z_{id}),\ldots,w_{k_n+l}(z_{id})).
\end{eqnarray*} 
The new definition of $\bar{w}$ allows us to easily enumerate the spline basis components for all $d$ nonlinear variables. Then the subgradient $s\left(\beta,\gamma\right) = \left(s_0(\beta,\gamma),\ldots,s_{p+dL_n}(\beta,\gamma)\right)$ of the corresponding objective function is given by
\begin{eqnarray*}
s_j(\vbeta, \vgamma)
&=& \frac{\tau}{n}\sumi \frac{R_i}{\tilde{\pi}_i}x_{ij}I(Y_i-\vx_i^\top(\vbeta_1,\mathbf{0}_{p_n-q_n}) - \vW(\vz_i)^\top\vgamma > 0)\\
&& + \frac{1-\tau}{n} \sumi \frac{R_i}{\tilde{\pi}_i} x_{ij}I(Y_i-\vx_i^\top(\vbeta_1,\mathbf{0}_{p_n-q_n}) - \vW(\vz_i)^\top\vgamma < 0) 
\\ &&- \frac{1}{n} \sumi \frac{R_i}{\tilde{\pi}_i} x_{ij}a_i \,\,\, \mbox{ for } 1 \leq j \leq p_n, \\
s_j(\vbeta, \vgamma) &=& \frac{\tau}{n}\sumi \frac{R_i}{\tilde{\pi}_i} \bar{w}_{j-(p_n+1)}(z_i)I(Y_i-\vx_i^\top(\vbeta_1,\mathbf{0}_{p_n-q_n}) - \vW(\vz_i)^\top\vgamma > 0) \\
&&+ \frac{1-\tau}{n} \sumi \frac{R_i}{\tilde{\pi}_i} \bar{w}_{j-(p_n+1)}(z_i)I(Y_i-\vx_i^\top(\vbeta_1,\mathbf{0}_{p_n-q_n}) - \vW(\vz_i)^\top\vgamma < 0) \\ &&- \frac{1}{n} \sumi \frac{R_i}{\tilde{\pi}_i} \bar{w}_{j-(p_n+1)}(z_i)a_i \,\,\, \mbox{ for } p_n + 1 \leq j \leq p_n+J_n,
\end{eqnarray*}
where $a_i=0$ if $Y_i-\vx_i^\top(\vbeta_1,\mathbf{0}_{p-q}) - \vW(\vz_i)^\top\vgamma \neq 0$, and $a_i \in [\tau-1,\tau]$ otherwise. For ease of notation in this proof let $(\hat{\beta},\hat{\gamma})$ represent the oracle estimator from (\ref{weightedOracle}). Following the proof of Theorem 2.4 from Wang, Wu and Li (2012) it is sufficient to show that with probability approaching one
\begin{eqnarray}
s_j\left(\hat{\vbeta}, \hat{\vgamma}\right) &=&0, \,\, j=1,\ldots,q_n \mbox{ or } \ j = p_n+1,\ldots,p_n+dL_n, \label{lem_wt_sub_1}\\
\left|\hat{\vbeta}_j\right| &\geq& (a+1/2)\lambda, \,\, j=1,\ldots,q_n,  \label{lem_wt_sub_2} \\
\left|s_j\left(\hat{\vbeta},\hat{\vgamma}\right)\right| &\leq& \lambda, \ j=q_n+1,\ldots,p_n. \label{lem_wt_sub_3}
\end{eqnarray}
Convex optimization theory immediately provides (\ref{lem_wt_sub_1}) holds, while (\ref{lem_wt_sub_2}) holds from the assumption that $n^{-1/2}q_n=o(\lambda)$, $\sqrt{q/n}$ consistency of $\hat{\vbeta}$ as stated in Theorem \ref{o_high_d_n} and Condition \ref{cond_small_sig} for a lower bound on the smallest true linear signal. Let $\mathcal{D}=\{i: Y_i - \xia^\top\hat{\vbeta}_1 - \vW(\vz_i)^\top\hat{\vgamma}  = 0\}$, then
for $j=q_n+1,\ldots,p_n$
\begin{equation*}
s_j\left(\hat{\vbeta},\hat{\vgamma}\right) = \frac{1}{n} \sumi \frac{R_i}{\tilde{\pi}_i} x_{ij} \left[I\left(Y_i - \xia^\top\hat{\vbeta}_1 - \vW(\vz_i)^\top\hat{\vgamma} \leq 0\right) - \tau \right]
 - \frac{1}{n} \sum_{i \in \mathcal{D}} \frac{R_i}{\tilde{\pi}_i} x_{ij}(a_i^* + (1-\tau)), 
\end{equation*}
where $a_i^* \in [\tau-1,\tau]$ such that $s_j(\hat{\vbeta},\hat{\vgamma})=0$ when $i \in \mathcal{D}$. With probability one (Section
2.2, Koenker, 2005), $|\mathcal{D}| = q_n+J_n$. Therefore by Conditions \ref{cond_g_h}, \ref{xmar_cond_alpha} and \ref{xmar_cond_eta} and the rate of $\lambda$ stated in Theorem \ref{thm_weighted_scad_n}, 
$
n^{-1}\sum_{i \in \mathcal{D}} x_{ij}(a_i^* + (1-\tau)) = {\rm O}_{\rm p}\left((q_n+dk_n+1)n^{-1}\right)={\rm o}_{\rm p}(\lambda).
$
Let $M_n$ denote the event that $\underset{i}{\max} \left| \tilde{\pi}_i-\pi_{i0} \right| \leq C\left(h^b+(\ln n / h^sn)^{1/2}\right)$, for some positive constant $C$. Then by assumptions of Theorem 6 it is sufficient to show

\begin{equation*}
\Pr\left( \underset{q_n+1 \leq j \leq p_n}{\max} \left|\frac{1}{n} \sumi \frac{R_i}{\tilde{\pi}_i} x_{ij}\left[ I(Y_i - \vx_{A_i}^\top\hat{\vbeta}_1 - \hat{g}(\vz_i) \leq 0) - \tau \right] \right|  > \lambda \middle \vert M_n \right) \rightarrow 0 \,\,\, \forall j.
\end{equation*}

Proof of Lemma 1 (3.5) from Sherwood and Wang (2016) can be modified to show  
\begin{equation*}
\underset{q_n+1 \leq j \leq p_n}{\max} \left| \frac{1}{n} \sumi \frac{R_i}{\pi_{i0}} x_{ij}\left[ I(Y_i - \xia^\top\hat{\vbeta} - \hat{g}(\vz_i) \leq 0) - \tau \right]\right| = o(\lambda).
\end{equation*}
Recall Condition 7 provides an upper bound of $|x_{ij}|$. Given $M_n$ holds Condition 4 and Condition 6 can be combined to derive an upper bound for $\underset{i}{\max}|\tilde{\pi}_i|$. Under the condition of $M_n$ holding then  
\begin{eqnarray*}
&& \underset{q_n+1 \leq j \leq p_n}{\max} \left|\frac{1}{n} \sumi R_i \left(\frac{1}{\tilde{\pi}_i}-\frac{1}{\pi_{i0}}\right) x_{ij} \left[ I(Y_i - \xia^\top\hat{\vbeta} - \hat{g}(\vz_i) \leq 0) - \tau \right]\right| \\
&\leq& C \left( \underset{i,j}{\max} |x_{ij}| \right) \left(\underset{i}{\max} \left| \frac{\tilde{\pi}_i-\pi_{i0}}{\tilde{\pi}_i\pi_{i0}} \right| \right) \leq C \underset{i}{\max} \left| \tilde{\pi}_i-\pi_{i0} \right| \\
&\leq& C \left(h^b+(\ln n / h^sn)^{1/2}\right) = o(\lambda).
\end{eqnarray*}
With the final rate coming from the assumption that $\left(h^b+(\log n /h^sn)^{1/2}\right)=o(\lambda)$. Therefore (\ref{lem_wt_sub_1}), (\ref{lem_wt_sub_2}) and (\ref{lem_wt_sub_3}) hold, completing the proof. 
\end{proof}

\noindent{\bf Acknowledgments.} I thank the Editor, the AE and the anonymous referees
for their careful reading and constructive comments which have helped significantly improve this paper.

\

\noindent{\large\bf References}
\begin{description}
\item Chen, X., Wan, A. and Zhou, Y. (2015). Efficient quantile regression analysis with missing observations. {\it J. Amer. Statist. Assoc.} {\bf 110}, 723-741.
\item Cheng, P. (1995) A note on strong convergence rates in nonparametric regression. {\it Statist. Probab. Lett.} {\bf 24}, 357-364.
\item De Gooijer, J. and Zerom, D. (2003). On additive conditional quantiles with high-dimensional covariates. {\it J. Amer. Statist. Assoc.} {\bf 98}, 135-146.
\item Fan, J. and Li, R. (2001). Variable selection via nonconcave penalized likelihood and its oracle properties. 
{\it J. Amer. Statist. Assoc.} {\bf 96}, 1348-1360.
\item  Fan, J. and Song, R. (2010). Sure independence screening in generalized linear models with NP-dimensionality. {\it Ann. Stat.} {\bf 38}, 3567–3604.
\item He, X. and Shi, P. (1996). Bivariate tensor-product B-splines in a partly linear model. {\it J. Multivar. Anal.} {\bf 58}, 162-181.
\item He, X., Zhu, Z. and Fung, W. (2002). Estimation in a semiparametric model for longitudinal data with unspecified dependence structure. {\it Biometrika} {\bf 89}, 579-590.
\item Horowitz, J. and Lee, S. (2005). Nonparametric estimation of an additive quantile regression model. {\it J. Amer. Statist. Assoc.} {\bf 100}, 1238-1249.
\item Hosmer, D., Lemeshow, S. and May, S. {\it Applied Survival Analysis: Regression Modeling of Time to Event Data} Second Edition. Wiley, New York. 
\item Huang, J., Wei, F., and Ma, S. (2012). Semiparametric regression pursuit. {\it Stat. Sinica}, {\bf 22}, 1403-1426.
\item Klemela, J. (2013) regpro: Nonparametric Regression. R package version
0.1. Available at https://cran.r-project.org/web/packages/regpro/index.html.
\item Koenker, R. (2005). Quantile Regression. Cambridge University Press. 
\item Koenker, R. and Bassett, G. (1978). Regression quantiles. {\it Econometrica} {\bf 46}, 33-50. 
\item Lavergne, P. (2008). A Cauchy-Schwarz inequality for expectation of matrices. Working
Paper, Simon Fraser University.
\item Lian, H., Liang, H. and Ruppert, D. (2015). Separation of covariates into nonparametric and parametric parts in high-dimensional partially linear additive models. {\it Stat. Sinica}, {\bf 25}, 591-607. 
\item Liang, H., Wang, S., Robins, J. and Caroll, R. (2004). Estimation in partially linear models with missing covariates. 
{\it J. Amer. Statist. Assoc.} {\bf 99}, 357-367.
\item Lipsitz, S., Fitzmaurice, G., Molenberghs, G. and Zhao, L. (1997). Quantile regression methods for longitudinal data with drop-outs: application to CD4 cell counts of patients infected with the human immunodeficiency virus. {\it J. R. Stat. Soc. Series C} {\bf 46}, 463-476.
\item Liu, T. and Yuan, X. (2015). Weighted quantile regression with missing covariates using empirical likelihood. {\it Statistics}, 1-25.
\item Liu, X., Wang, L. and Liang, H. (2011). Estimation and variable selection for semiparametric additive partial linear models. {\it Stat. Sinica} {\bf 21}, 1225-1248.
\item Nadaraya, E. (1964). On estimating regression. {\it Theory Probab. Appl}, {\bf 10}, 186-190.
\item   Robins, J., Rotnitsky, A. Zhao, L. (1994). Estimation of regression coefficients when some regressors are not always observed. {\it J. Amer. Statist. Assoc.} {\bf 89}, 846-866.
\item Schwarz, G. (1978), Estimating the dimension of a model. {\it Ann. Stat.} {\bf 6}, 461–464.
\item Schumaker, L. (1981) Spline Functions: Basic Theory. Wiley: New York.
\item Sepanski, J., Knickerbocker, R. and Carroll, R. (1994), A semiparametric
correction for attenuation. {\it J. Amer. Statist. Assoc.} {\bf 89}, 1366–1373.
\item Sherwood, B. and Maidman, A. (2016) rqPen: Penalized Quantile Regression. R package version
1.4. Available at https://cran.r-project.org/web/packages/rqPen/index.html.
\item Sherwood, B. and Wang, L. (2016). Additive Partially Linear Quantile Regression in Ultra-high Dimension. {\it Ann. Stat.} {\bf 44}, 288-317.
\item  Sherwood, B., Wang, L. and Zhou, X. (2013). Weighted quantile regression for analyzing health care cost data with missing covariates.  {\it Stat. Med.} {\bf 32}, 4967-4979.
\item Stone, C. (1985). Additive regression and other nonparametric models. {\it Ann. Stat.} {\bf 13}, 689-706.
\item Tao, P. and An, L. (1997). Convex analysis approach to D.C. programming: theory, algorithms and applications. {\it Acta Math. Vietnam.} \textbf{22}, 289-355.
\item Tibshirani, R. (1996). Regression shrinkage and selection via the Lasso. {\it J. R. Stat. Soc., Series B} {\bf 58}, 267-288.
\item Tripathi, G. (1999). A matrix extension of the Cauchy-Schwarz inequality. {\it Economics Letters} {\bf 63}, 1-3. 
\item Wang, C., Wang, S., Zhao, L., and Ou, S. (1997). Weighted semiparametric
estimation in regression analysis with missing covariate data. {\it J. Amer. Statist. Assoc.} {\bf 92}, 512–525.
\item Wang, C.Y., Wang, S., Gutierrez, R. and Carroll, R.J. (1998) Local linear regression for generalized linear models with missing data. {\it Ann. Stat.} {\bf 26}, 1028-1050. 
\item Wang, H., Zhu, Z. and Zhou, J. (2009). Quantile regression in partially linear varying coefficient models. {\it Ann. Stat.} {\bf 37}, 3841-3866.
\item Wang, L., Wu, Y. and Li, R. (2012). Quantile regression for analyzing heterogeneity in ultra-high dimension. {\it J. Amer. Statist. Assoc.} {\bf 107}, 214-22.
\item Watson, G. (1964). Smooth regression analysis. {\it Sankhya Ser. A} {\bf 26}, 359-372. 
\item Wei, Y., Ma., Y. and Carroll, R. (2012). Multiple imputation in quantile regression. {\it Biometrika} {\bf 99}, 423-438.
\item Wei, Y. and Yang, Y. (2014). Quantile regression with covariates missing at random. {\it Stat Sinica} {\bf 24}, 1277-1299. 
\item Wu, Y. and Liu, Y. (2009). Variable selection in quantile regression. {\it Stat Sinica} {\bf 19}, 801-817.
\item Yi, G. and He, W. (2009). Median regression models for longitudinal data with dropouts. {\it Biometrics} {\bf 65}, 618-625.
\item Zhang, C. (2010).  Nearly unbiased variable selection under minimax concave penalty. {\it Ann. Stat.} {\bf 38}, 894-942.
\item Zhang, H., Cheng, G. and Liu, Y. (2011). Linear or nonlinear? Automatic structure discovery for partially linear models. {\it J. Amer. Statist. Assoc.}, {\bf 106}, 1099-1112.
\item Zhao, P. and Yu, B. (2006). On model selection consistency of lasso. {\it JMLR} {\bf 7}, 2541-2563.
\item Zou, H. and Li, R. (2008). One-step sparse estimates in nonconcave penalized likelihood models. {\it Ann. Stat.} {\bf 36}, 1509-1533.

\end{description}


\vskip .65cm
\noindent
Johns Hopkins University
\vskip 2pt
\noindent
E-mail: bsherwo2@jhu.edu
\vskip 2pt

\end{document}